\documentclass[twocolumn]{autart}
\usepackage{mathtools}
\usepackage{amssymb}
\usepackage{enumitem}
\usepackage{mathdots}
\usepackage{xfrac}
\usepackage{bm}
\usepackage{bbm}
\usepackage{array}
\usepackage{widetext}
\usepackage{algorithm}
\usepackage{algorithmicx}
\usepackage{algpseudocode}
\usepackage{comment}
\usepackage{pgfplots}
\usepackage{lipsum}
\usepackage{tikz}

\usetikzlibrary{arrows,calc,patterns}
\tikzstyle{Rect}=[draw=gray,line width=0.001pt,preaction={clip, postaction={pattern=north east lines, pattern color=gray,line width=0.1pt}}]
\tikzset{
	>=stealth',
	help lines/.style={dashed, thick},
	axis/.style={<->},
	important line/.style={thick},
	connection/.style={thick, dotted},
}

\definecolor{MatlabBlue}{rgb}    {0     , 0.4470, 0.7410}
\definecolor{MatlabRed}{rgb}     {0.8500, 0.3250, 0.0980}
\definecolor{MatlabYellow}{rgb}  {0.9290, 0.6940, 0.1250}
\definecolor{MatlabPurple}{rgb}  {0.4940, 0.1840, 0.5560}
\definecolor{MatlabGreen}{rgb}   {0.4660, 0.6740, 0.1880}
\definecolor{MatlabBabyBlue}{rgb}{0.3010, 0.7450, 0.9330}
\definecolor{MatlabGray}{rgb}{0.5, 0.5, 0.5}
\definecolor{MatlabLightGray}{rgb}{0.75, 0.75, 0.75}
\definecolor{MatlabBlack}{rgb}{0, 0, 0}
\definecolor{MatlabLightGray4}{rgb}{0.875, 0.875, 0.875}
\definecolor{MatlabLightGray3}{rgb}{0.85, 0.85, 0.85}
\definecolor{MatlabLightGray2}{rgb}{0.775, 0.775, 0.775}
\definecolor{MatlabLightGray1}{rgb}{0.7, 0.7, 0.7}
\definecolor{MatlabGray20}{rgb}{0.2, 0.2, 0.2}
\definecolor{MatlabGray30}{rgb}{0.3, 0.3, 0.3}
\definecolor{MatlabGray40}{rgb}{0.4, 0.4, 0.4}
\definecolor{MatlabGray50}{rgb}{0.5, 0.5, 0.5}
\definecolor{MatlabGray60}{rgb}{0.6, 0.6, 0.6}
\definecolor{MatlabGray70}{rgb}{0.7, 0.7, 0.7}
\definecolor{MatlabGray80}{rgb}{0.8, 0.8, 0.8}
\definecolor{MatlabGray85}{rgb}{0.85, 0.85, 0.85}
\definecolor{MatlabGray90}{rgb}{0.9, 0.9, 0.9}
\definecolor{dblue}{rgb}    {0.1098    0.1294    0.3608}

\newcommand{\hop}{\mathsf{H}}
\newtheorem{assumption}{Assumption}
\newtheorem{lemma}{Lemma}

\newtheorem{theorem}{Theorem}

\usepackage{etoolbox}

\makeatletter
\def\@xnamedef#1{\expandafter\protected@xdef\csname #1\endcsname}
\def\no@harm{} 
\def\ead@au#1{\protected@edef\@ead@au{#1}}
\patchcmd\runningauthor@fmt{\global\edef}{\protected@xdef}{}{}
\patchcmd\runningauthor@fmt{\global\edef}{\protected@xdef}{}{}
\patchcmd\author@fmt{\edef}{\protected@edef}{}{}
\patchcmd\add@xtok{\xdef}{\protected@xdef}{}{}
\makeatother

\usepackage{eso-pic}
\AddToShipoutPictureBG*{%
	\AtPageUpperLeft{%
		\setlength\unitlength{1in}%
		\hspace*{\dimexpr0.5\paperwidth\relax}
		\makebox(8,-1.3)[c]{
			\begin{tabular}{c c}
				Rodrigo A. Gonz\'alez \emph{et al.},
				Statistically Optimal Structured Additive MIMO Continuous-time System Identification. \\
				uploaded to arXiv on May 20th, 2025. \\
\end{tabular}}}}
                                           
\usepackage{color}
\pgfplotsset{compat=1.18}

\begin{document}
	\begin{frontmatter}
		\title{Statistically Optimal Structured Additive MIMO Continuous-time System Identification\thanksref{footnoteinfo}} 
		
		\thanks[footnoteinfo]{The material in this paper was not presented at any IFAC meeting. Corresponding author: Rodrigo A. Gonz\'alez.}
		
		\author[TUE]{Rodrigo A. Gonz\'alez}\ead{r.a.gonzalez@tue.nl},    
		\author[TUE]{Maarten van der Hulst}\ead{m.v.d.hulst@tue.nl}, %
            \author[TUE]{Koen Classens}\ead{k.h.j.classens@tue.nl}, %
            \author[TUE,Delft]{Tom Oomen}\ead{t.a.e.oomen@tue.nl}   
		\address[TUE]{Control Systems Technology Section, Department of Mechanical Engineering, Eindhoven University of Technology, Eindhoven, The Netherlands.} 
            \address[Delft]{Delft Center for Systems and Control, Delft University of Technology, Delft, The Netherlands.}  
		
		\vspace{-0.2cm}
		\begin{abstract}
                Many applications in mechanical, acoustic, and electronic engineering require estimating complex dynamical models, often represented as additive multi-input multi-output (MIMO) transfer functions with structural constraints. This paper introduces a two-stage procedure for estimating structured additive MIMO models, where structural constraints are enforced through a weighted nonlinear least-squares projection of the parameter vector initially estimated using a recently developed refined instrumental variables algorithm. The proposed approach is shown to be consistent and asymptotically efficient in open-loop scenarios. In closed-loop settings, it remains consistent despite potential noise model misspecification and achieves minimum covariance among all instrumental variable estimators. Extensive simulations are performed to validate the theoretical findings, and to show the efficacy of the proposed approach.
                \vspace{-0.3cm}
            \end{abstract}
		\end{frontmatter}
	\section{Introduction}
\vspace{-0.2cm}

In system identification, the goal is to develop a mathematical model of a dynamical system using both prior knowledge and available data. Depending on the sampling and modeling approach, these models can be formulated in discrete time \cite{Ljung1999SystemUser,Soderstrom2001SystemIdentification} or in continuous time \cite{Garnier2008book}, with the latter often directly reflecting the underlying physical laws via ordinary or partial differential equations.

System identification methods differ significantly when open-loop or closed-loop data is used. For systems operating in closed loop, additional constraints, including stricter conditions on identifiability \cite{ljung1974identification}, persistence of excitation, and noise modeling, must be satisfied to achieve consistent and asymptotically efficient estimators \cite{VanDenHof1998AnnualIdentification}. Direct methods for closed-loop identification, which ignore the feedback connection to identify an open-loop model solely from input and output data, require a careful selection of the noise model to mitigate bias \cite{forssell1999closed}. An alternative to this approach is to employ instrumental variables \cite{soderstrom1987instrumental}, which have the potential to yield asymptotically unbiased estimators even when the noise model is misspecified \cite{Gilson2005InstrumentalIdentification}. The aforementioned methods, initially conceived for discrete-time systems, have also been developed for continuous-time system identification \cite{Gilson2004InstrumentalClosed-loop,gilson2008instrumental}.

For either open or closed-loop system identification, extending the aforementioned methods from single-input single-output (SISO) to multi-input multi-output (MIMO) models introduces further challenges in numerical robustness, model structure flexibility, and theoretical guarantees. Optimization-based methods such as the prediction error method or maximum likelihood estimation can be applied directly to MIMO systems, although they often lead to highly non-convex problems that lack a unique global optimum and are sensitive to initialization. Various strategies have been proposed to address these issues, including solving the maximum likelihood problem via the Expectation-Maximization (EM) algorithm for open-loop scenarios \cite{gibson2005robust}, employing subspace methods \cite{ljung1996subspace,van2013closed}, and, more recently, using the weighted null space fitting method \cite{he2024weighted}. In continuous-time system identification, these methods may lead to overparameterization, fail to directly estimate physical parameters \cite{vayssettes2015structured}, lack proven statistical efficiency, or, as with the weighted null space fitting method, currently lack a continuous-time counterpart. Other multi-step and iterative approaches, such as the Box-Jenkins Steiglitz-McBride algorithm \cite{zhu2016box} and refined instrumental variables \cite{Young1980RefinedAnalysis}, were initially designed for discrete-time SISO or multi-input single-output (MISO) models with common denominator structures. Extensions of the instrumental variable methods to more general model structures have been proposed for continuous-time MISO systems with differing denominator polynomials \cite{Garnier2007AnSystems} and for discrete-time matrix fraction descriptions \cite{Akroum2009ExtendingModels,Blom2010MultivariableRegression}. However, these methods tend to be less parsimonious and do not readily accommodate the extra structural constraints often needed in more complex applications.

Recognizing these limitations, recent developments have focused on alternative parametrizations that naturally incorporate more structure into the model. In many applications, imposing tailored parametrizations on structured MIMO systems leads to more parsimonious models that offer deeper physical insight. One particularly useful parametrization is that of additive MIMO systems, where the overall system is expressed as the sum of simpler MIMO subsystems. Applications that use this model structure can be found in acoustics \cite{jian2022acoustic}, mechatronics \cite{Voorhoeve2021IdentifyingStage,Shirvani2023LinearData,tacx2024identification}, thermal systems \cite{zhu2008robust}, civil \cite{zhang2019variational}, electronic \cite{lange2020broadband} and aerospace engineering \cite{Vayssettes2014Frequency-domainTests}, and fault detection \cite{Classens2024RecursiveSystems}. Initial methods for identifying additive SISO systems in both open- and closed-loop settings were introduced in \cite{Gonzalez2024IdentificationClosed-loop}, and later extended to additive MIMO systems in \cite{vanderHulst2024additivecdc}. Beyond the MIMO additive system representation, it is often beneficial to impose additional structure on its subsystem components. For instance, constraining the numerator matrices within each additive subsystem to have reduced rank can facilitate modal analysis \cite{Gawronski2004AdvancedStructures,vanderHulst2025FrequencyStage}. State-space identification methods incorporating null-space and gradient-based optimization have been applied to enhance the parsimony of black-box models \cite{mercere2014identification}, although its specificity to deal with additive systems is lacking in the first identification step.

Despite these recent advances, two key challenges remain. First, the refined instrumental variable method proposed in \cite{vanderHulst2024additivecdc} lacks a comprehensive bias and variance analysis, which is critical for the applicability of the method in general identification setups. Second, there is currently no statistically optimal framework for incorporating structural constraints, such as rank-reduced numerators, into the additive continuous-time MIMO setting using time-domain data. This paper addresses both of these gaps. The main contributions are as follows:

\begin{enumerate}[label=C\arabic*]
\label{contribution1}
	\item We provide a comprehensive statistical analysis of a refined instrumental variables method for estimating additive MIMO models from time-domain open-and closed-loop data. Specifically, we show that the method proposed in \cite{vanderHulst2024additivecdc} is generically consistent under mild conditions without requiring parametric noise modeling, and we derive its asymptotic covariance to show that it achieves asymptotic efficiency for open-loop data. We also quantify the loss of statistical efficiency in closed-loop operation, which arises because the instrument matrix omits noise components to relax consistency conditions on the noise model.
\label{contribution2}
    \item We propose a two-stage procedure for estimating structured additive MIMO models, where additional structure is enforced through a weighted nonlinear least-squares projection of the parameter vector estimated using refined instrumental variables. We prove that the optimal statistical properties established in Contribution 1 extend to this framework, and we formally link this approach to the indirect prediction error method \cite{Soderstrom1991AnIdentification} for open-loop operation.
    \label{contribution3}
	\item We support our theoretical findings with extensive numerical simulations that verify the statistical properties of the proposed identification methods.
\end{enumerate}
The remainder of this paper is structured as follows. Section \ref{sec:system} introduces the additive system and model structure, and presents the identification problem. The refined instrumental variable approach for additive MIMO systems is outlined in Section \ref{sec:riv}, and a comprehensive statistical analysis of the method for open and closed loop is presented in Section \ref{sec:statistical}. The second step of the refined instrumental variable method for structured additive systems is detailed in Section \ref{sec:ipem}. Extensive simulations that verify the theoretical results and applicability of the methods are provided in Section \ref{sec:simulations}, and conclusions are drawn in Section \ref{sec:conclusions}. Proofs of supplementary technical results are given in the Appendix.

\textit{Notation}: All matrices, vectors, and vector-valued signals are written in bold, and vectors are column vectors, unless transposed. The operator $\sigma_{\min }(\cdot)$ represents the minimum singular value of a matrix, $p$ is the Heaviside operator, i.e., $p \mathbf{u}(t)=\frac{\mathrm{d}}{\mathrm{d} t}\mathbf{u}(t)$ for $t\in \mathbb{R}$, and $q$ is the forward shift operator $q\mathbf{u}(kh) = \mathbf{u}([k+1]h)$ for $k\in \mathbb{Z}$. The expected value of a quasi-stationary signal $\mathbf{x}(t_k)$ is defined as $\overline{\mathbb{E}}\{\mathbf{x}(t_k)\}=\lim_{N\to\infty}N^{-1}\sum_{k=1}^N \mathbb{E}\{\mathbf{x}(t_k)\}$. The notation $x_N = o_p(a_N)$ indicates that $x_N/a_N$ converges to zero in probability as $N\to\infty$, $y_N=O_p(b_N)$ means that $y_N/b_N$ is bounded in probability as $N\to\infty$.

\vspace{-0.2cm}
\section{System and model setup}
\vspace{-0.2cm}
\label{sec:system}
Consider the multi-input multi-output (MIMO), linear and time-invariant (LTI), causal continuous-time system with $n_{\mathrm{u}}$ inputs and $n_{\mathrm{y}}$ outputs in additive form
\begin{equation} \label{eq: CT - plant model}
\mathbf{x}(t) = \sum_{i=1}^K\mathbf{G}^*_i(p)\mathbf{u}(t),
\end{equation}
where the input signal is denoted by $\mathbf{u}(t) \in \mathbb{R}^{n_{\mathrm{u}}}$ and  $\mathbf{x}(t) \in \mathbb{R}^{n_{\mathrm{y}}}$ is the unobserved noise-free plant output. Each true subsystem $\mathbf{G}_i^*(p)$ can be expressed as
\begin{equation} \label{eq: CT - subsystem model}
\mathbf{G}^*_i(p) = \dfrac{1}{A^*_i(p)}\mathbf{B}^*_i(p),
\end{equation}
where the true scalar denominator polynomial $A_i^*(p)$ and the matrix numerator polynomial $\mathbf{B}_i^*(p)$ are defined as
\begin{align} 
A^*_i(p) &= 1 + a^*_{i,1} p + \dots + a^*_{i,n_i}p^{n_i}, \label{eq: CT - plant polynomial A}\\
\mathbf{B}^*_i(p) &= \mathbf{B}^*_{i,0} + \mathbf{B}^*_{i,1} p + \dots + \mathbf{B}^*_{i,m_i}p^{m_i}. \label{eq: CT - plant polynomial B}
\end{align}
The $A_i^*(p)$ polynomials are assumed stable, and jointly coprime, i.e., they do not share roots. In addition, no complex number $s$ simultaneously satisfies $A_i^*(s)=0$ and $\mathbf{B}_i^*(s)=\mathbf{0}$, which implies the coprimeness of the numerator and denominator if the system is SISO. For obtaining a unique characterization of $\left\{\mathbf{G}_i^*(p)\right\}_{i=1}^K$, we assume that at most one subsystem $\mathbf{G}_i^*(p)$ is biproper. The denominator and numerator polynomials are jointly described by the true parameter vector
\begin{align} \label{eq: CT - parameter vector}
\boldsymbol{\beta}^* = \left[ \boldsymbol{\theta}^{*\top}_1,\ \dots,\ \boldsymbol{\theta}^{*\top}_K \right]^{\top},
\end{align}
where each $i$th subsystem is described by
\begin{align} \label{eq: CT - parameter vector subsystem}
\boldsymbol{\theta}^*_i \hspace{-0.07cm}= \hspace{-0.1cm}\left[ a^{*}_{i,\hspace{-0.02cm}1}\hspace{-0.02cm}, \dots\hspace{-0.02cm},\hspace{-0.02cm} a^{*}_{i,n_i}\hspace{-0.02cm},\hspace{-0.02cm} \operatorname{vec}\{\hspace{-0.02cm}\mathbf{B}^{*}_{i,0}\hspace{-0.01cm}\}^{\hspace{-0.04cm}\top}\hspace{-0.05cm}, \dots\hspace{-0.02cm}, \operatorname{vec}\{\hspace{-0.02cm}\mathbf{B}^{*}_{i,m_i}\hspace{-0.03cm}\}^{\hspace{-0.02cm}\top}\hspace{-0.02cm}\right]^{\hspace{-0.03cm}\top}\hspace{-0.09cm}.
\end{align}
A noisy measurement of the output is retrieved at every time instant $t=t_k,\ k=1, \ldots, N$, where $\left\{t_k\right\}_{k=1}^N$ are equidistant in time. That is,
\begin{equation} \label{eq: CT - output observation}
\mathbf{y}(t_k) 
= \mathbf{x}(t_k) + \mathbf{v}(t_k),
\end{equation}
where $\mathbf{v}\left(t_k\right)$ is a zero-mean discrete-time stationary random process of unknown positive definite covariance $\bm{\Sigma}_0$.

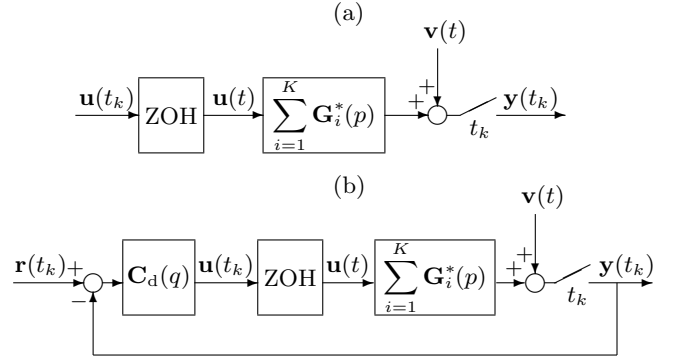
\begin{figure}
	\setlength{\unitlength}{0.092in} 
	\centering 
	\begin{picture}(37,20) 
		\put(18,19){\small{(a)}}
		\put(3.5,13.6){\vector(1,0){3.6}}
		\put(3.7,14.2){\footnotesize{$\mathbf{u}(t_k)$}}
		\put(7.1,11.5){\framebox(3.6,4.2){\footnotesize{ZOH}}}
		\put(10.7,13.6){\vector(1,0){3.4}}
		\put(11.2,14.2){\footnotesize{$\mathbf{u}(t)$}}
		\put(14.1,11.5){\framebox(6.8,4.2){\footnotesize{$\displaystyle\sum_{i=1}^K \mathbf{G}_i^*\hspace{-0.02cm}(p)$}}}
		\put(20.9,13.6){\vector(1,0){2.5}}
		\put(22.2,14.1){\scriptsize{+}}
		\put(23.1,18){\footnotesize{$\mathbf{v}(t)$}}
		\put(23.9,13.6){\circle{1}}
		\put(23.9,17.4){\vector(0,-1){3.3}}
		\put(22.8,14.9) {\scriptsize{+}}
		\put(24.4,13.6){\line(1,0){0.8}}
		\put(25.2,13.6){\line(2,1){2}}
		\put(25.7,12.5){\footnotesize{$t_k$}}	
		\put(27.2,13.6){\vector(1,0){3.9}}
		\put(27.6,14.2){\footnotesize{$\mathbf{y}(t_k)$}}	

		\put(0,4.1){\vector(1,0){4}}
		\put(0.1,4.7){\footnotesize{$\mathbf{r}(t_k)$}}
		\put(3,4.6){\scriptsize{+}}
		\put(18,9.2){\small{(b)}}
		\put(3.2,2.7){\scriptsize{$-$}}
		\put(4.5,0){\vector(0,1){3.6}}
		\put(4.5,0){\line(1,0){29.5}}
		\put(4.5,4.1){\circle{1}}
		\put(5,4.1){\vector(1,0){1.2}}
		\put(6.2,2.3){\framebox(4,4.2){\small{$\mathbf{C}_\textnormal{d}(q)$}}}
		\put(10.2,4.1){\vector(1,0){3.6}}
		\put(10.4,4.7){\footnotesize{$\mathbf{u}(t_k)$}}
		\put(13.8,2.3){\framebox(3.6,4.2){\footnotesize{ZOH}}}
		\put(17.4,4.1){\vector(1,0){2.9}}
		\put(17.6,4.7){\footnotesize{$\mathbf{u}(t)$}}
		\put(20.4,2.3){\framebox(6.7,4.2){\footnotesize{$\displaystyle\sum_{i=1}^K \mathbf{G}_i^*\hspace{-0.02cm}(p)$}}}
		\put(27.2,4.1){\vector(1,0){1.6}}
		\put(27.7,4.6){\scriptsize{+}}
		\put(28.6,8.6){\footnotesize{$\mathbf{v}(t)$}}
		\put(29.4,4.1){\circle{1}}
		\put(29.4,8){\vector(0,-1){3.4}}
		\put(28.3,5.4) {\scriptsize{+}}
		\put(29.9,4.1){\line(1,0){0.7}}
		\put(30.5,4.1){\line(2,1){2}}
		\put(31.2,3){\footnotesize{$t_k$}}	
		\put(34,0){\line(0,1){4.1}}
		\put(32.4,4.1){\vector(1,0){3.6}}
		\put(33,4.7){\footnotesize{$\mathbf{y}(t_k)$}}	
	\end{picture}
	\caption{Block diagrams for the open (a) and closed-loop (b) system settings.}
	\label{fig_setups}
\end{figure}

Two scenarios are considered for the generation of the sampled input $\mathbf{u}(t_k)$, as presented in Figure \ref{fig_setups}. In the \textit{open-loop} case, the input is assumed to be statistically independent of the disturbance $\mathbf{v}(t_k)$. In contrast, the \textit{closed-loop} case assumes that the input is a function of the measured output $\mathbf{y}(t_k)$ and reference signal $\mathbf{r}(t_k)\in\mathbb{R}^{n_{\textnormal{y}}}$, and is thus correlated with the disturbance signal. More precisely, the closed-loop scheme results in the following description for $\mathbf{u}(t_k)$:
\begin{equation}
\label{uclosedloop}
    \mathbf{u}(t_k) = \mathbf{S}_{uo}^*(q)\left(\mathbf{r}(t_k)-\mathbf{v}(t_k)\right),
\end{equation}
where the control sensitivity function is given by $\mathbf{S}_{uo}^*(q)=\mathbf{C}_{\textnormal{d}}(q)[\mathbf{I}_{n_{\textnormal{y}}}+\mathbf{G}_\textnormal{d}^*(q)\mathbf{C}_{\textnormal{d}}(q)]^{-1}$, with $\mathbf{C}_{\textnormal{d}}(q)$ being a non-zero discrete-time transfer function representing the controller, and with $\mathbf{G}_\textnormal{d}^*(q)$ being the zero-order hold (ZOH) discrete-time equivalent of the total plant $\mathbf{G}^*(p) = \sum_{i=1}^K \mathbf{G}_i^*(p)$. For either scenario, the continuous-time input to the system is assumed to be constant between samples, although extensions to arbitrary input intersample behavior are possible \cite{gonzalez2021srivc}.

The additive structure in \eqref{eq: CT - plant model} is particularly beneficial for promoting parsimony of the resulting model \cite{Gonzalez2023ParsimoniousApproach}, and can lead to more tractable iterative algorithms for parametric identification of stiff and high-order systems \cite{vanderHulst2024additivecdc}. However, several applications may require imposing additional constraints on the parameter vector $\bm{\beta}^*$, reducing its effective dimension. Thus, in addition to the additive model structure of the system in \eqref{eq: CT - plant model}-\eqref{eq: CT - subsystem model}, we assume the existence of a known function that links the parameter vector $\bm{\beta}$ of the additive MIMO model to a parameter vector $\bm{\rho}$ of smaller dimension, pertaining to a \textit{structured} additive MIMO model. As an example of the applicability of this framework, we consider the modal decomposition \cite{Gawronski2004AdvancedStructures} of multivariable mechanical systems, described by
\begin{equation}
\label{multivariablemodal}
    \mathbf{G}^*(p) = \sum_{i=1}^K \frac{\bm{\psi}_{l,i}\bm{\psi}_{r,i}^\top}{1+2(\xi_i/\omega_i)p+p^2/\omega_i^2},
\end{equation}
where $\omega_i$ and $\xi_i$ are the natural frequency and damping ratio of the $i$th submodel, respectively, and $\bm{\psi}_{l,i}$ and $\bm{\psi}_{r,i}$ are its corresponding left and right mode-shape vectors. In this example, the function $\mathbf{f}$ links the parameter vector $\bm{\beta}$ to the vector composed by the mode shapes and natural frequencies of each mode:
\begin{equation}
    \bm{\rho} = \left[ \boldsymbol{\rho}^{\top}_1,\ \dots,\ \boldsymbol{\rho}^{\top}_K \right]^{\top}, \quad \bm{\rho}_i = \left[ \xi_i,\omega_i, \bm{\psi}_{l,i}^\top, \bm{\psi}_{r,i}^\top\right]^{\top}. \notag    
\end{equation}
The aim of this paper is twofold:
\begin{enumerate}
    \item To develop and analyze the statistical properties of a method that allows for the direct identification of $\bm{\beta}^*$ using the input-output dataset $\{\mathbf{u}(t_k), \mathbf{y}(t_k)\}_{k=1}^N$ for the open-loop case, and $\{\mathbf{r}(t_k), \mathbf{u}(t_k), \mathbf{y}(t_k)\}_{k=1}^N$ for the closed-loop case (Contribution C1). 
    \item To use the derived statistical properties to propose an asymptotically optimal method for identifying structured additive continuous-time MIMO systems, and demonstrate its effectiveness (Contributions C2 and C3).
\end{enumerate}

\section{Refined instrumental variable method for MIMO additive systems}
\label{sec:riv}
In this section we present a method for identifying continuous-time MIMO systems in additive form, and establish relationships between the proposed method and refined instrumental variable approaches \cite{Young1980RefinedAnalysis}. The identification problem is formulated based on the residual output signal, which is computed as the difference between the measured system output $\mathbf{y}(t_k)$ and the simulated model output 
\begin{equation} \label{eq: CT - residual function}
\begin{aligned}
\boldsymbol{\varepsilon}(t_k,\boldsymbol{\beta}) & =\mathbf{y}(t_k) -  \sum_{i=1}^K\mathbf{G}_i(p,\boldsymbol{\theta}_i)\mathbf{u}(t_k), 
\end{aligned}
\end{equation}
where $\mathbf{G}_i(p,\boldsymbol{\theta}_i)$ is a transfer function parametrized as in \eqref{eq: CT - subsystem model} representing the model of the $i$th subsystem.

\begin{rem}
\label{rem1}
In \eqref{eq: CT - residual function} a mixed notation of continuous-time filters with sampled signals is introduced, previously considered in \cite{Garnier2008book}. If $\mathbf{G}(p)$ is a causal continuous-time filter and $\mathbf{u}(t_k)$ is a sampled signal, then $\mathbf{G}(p)\mathbf{u}(t_k)$ implies that the signal is first interpolated using a specified intersample behavior (i.e., ZOH or piecewise constant between samples) before being filtered by $\mathbf{G}(p)$. The continuous-time output is then sampled at $t=t_k$. The notation $\mathbf{u}^\top (t_k)\mathbf{H}(p)$ can be interpreted similarly.
\end{rem}

For both open and closed-loop scenarios a unified estimator $\hat{\bm{\beta}}_N$ is considered, obtained as an element of the solution set of the correlation equations
\begin{equation} \label{correlation}
\hat{\bm{\beta}}_N \in \left\{\bm{\beta} \,\bigg|\, \sum^N_{k=1} \hat{\mathbf{\Phi}}(t_k, \bm{\beta}) \mathbf{\Sigma}^{-1}(\bm{\beta})\boldsymbol{\varepsilon}(t_k,\bm{\beta})=\boldsymbol{0}\right\}.
\end{equation}
That is, $\hat{\bm{\beta}}_N$ is designed such that the residual output, after whitening by the inverse of the estimated covariance matrix $\bm{\Sigma}(\bm{\beta})$, is uncorrelated with the so called instrument matrix $\hat{\mathbf{\Phi}}\left(t_k, \boldsymbol{\beta}\right)$. Such an instrument matrix can be interpreted as a noiseless version of the Jacobian of the residual output with respect to $\bm{\beta}$, and is composed by vertically stacking the matrices $\hat{\mathbf{\Phi}}_i\left(t_k, \boldsymbol{\beta}\right)$ for $i=1,\dots,K$:
\begin{align} \label{instrumentmatrix}
\hat{\mathbf{\Phi}}_{i}\left(t_k, \boldsymbol{\beta}\right) =&\ \bigg[ \dfrac{-p\mathbf{B}_i(p)}{A_i^2(p)}\mathbf{z}(t_k), \dots,\ \dfrac{-p^{n_i}\mathbf{B}_i(p)}{A_i^2(p)}\mathbf{z}(t_k),\notag \\
&\dfrac{1}{A_i(p)}\mathbf{Z}^{\top}(t_k),\ \dots, \dfrac{p^{m_i}}{A_i(p)}\mathbf{Z}^{\top}(t_k)  \bigg]^{\top},
\end{align}
where the polynomials $A_i(p)$ and $\mathbf{B}_i(p)$ depend on $\bm{\beta}$, and $\mathbf{Z}(t_k)=\mathbf{z}(t_k) \otimes \mathbf{I}_{n_{\textnormal{y}}}$. The vector $\mathbf{z}(t_k)$ is equal to $\mathbf{u}(t_k)$ in the open-loop case, and equal to the noiseless input $\mathbf{S}_{uo}(q,\bm{\beta})\mathbf{r}(t_k)$ for the closed-loop case, in which the control sensitivity function $\mathbf{S}_{uo}(q,\bm{\beta})$ is constructed from the total model $\sum_{i=1}^K \mathbf{G}_i(p,\bm{\theta}_i)$. Note that the correlation equations in \eqref{correlation} do not, in general, yield a unique estimator $\hat{\bm{\beta}}_N$, as swapping pairs of subsystems alters the arrangement of parameters while still satisfying \eqref{correlation}. 

The main computational task, covered next, is the development of an iterative procedure to find the estimator $\hat{\boldsymbol{\beta}}_N$ that satisfies the condition \eqref{correlation} given data. To find a solution to the correlation condition in \eqref{correlation}, a pseudolinear regression algorithm \cite{solo1979thesis} is introduced. In this context, we note that the residual can be written as a pseudolinear regression for each $\bm{\theta}_i$ as follows:
\begin{equation}
\label{pseudolinear}
    \boldsymbol{\varepsilon}\left(t_k, \boldsymbol{\beta}\right) =   \tilde{\mathbf{y}}_{i,f}\left(t_k, \boldsymbol{\beta}\right) -\boldsymbol{\Phi}_i^{\top}\left(t_k, \boldsymbol{\beta}\right) \boldsymbol{\theta}_i,
\end{equation}
where the filtered residual output of the $i$th subsystem is defined as $\mathbf{y}_{i,f}(t_k, \boldsymbol{\beta})=A_i^{-1}(p)\tilde{\mathbf{y}}_{i}(t_k, \boldsymbol{\beta})$, with
\begin{equation} \label{eq: CT - residual output subsystem}
\tilde{\mathbf{y}}_{i}\left(t_k, \boldsymbol{\beta}\right) = \mathbf{y}(t_k) -\hspace{-0.05cm} \sum_{\substack{\ell=1,\dots,K \\ \ell \neq i}}\frac{1}{A_\ell(p)}\mathbf{B}_\ell(p)\mathbf{u}(t_k).
\end{equation}
The corresponding regressor matrix is given by
\begin{align} \label{eq: CT - regressor matrix subsystem}
\mathbf{\Phi}_{i}\left(t_k, \boldsymbol{\beta}\right) =&\ \left[ \dfrac{-p}{A_i(p)}\tilde{\mathbf{y}}_i(t_k),\dots,\ \dfrac{-p^{n_i}}{A_i(p)}\tilde{\mathbf{y}}_i(t_k) \right. \notag \\
&\left. \dfrac{1}{A_i(p)}\mathbf{U}^{\hspace{-0.03cm}\top}(t_k),\dots,\dfrac{p^{m_i}}{A_i(p)}\mathbf{U}^{\hspace{-0.03cm}\top}(t_k)  \right]^{\top},
\end{align}
with $\mathbf{U}(t_k) = \mathbf{u}(t_k)\otimes \mathbf{I}_{n_{\textnormal{y}}}$. Replacing the pseudolinear regression \eqref{pseudolinear} into \eqref{correlation} for each $i$ and stacking by columns allows for the condition in \eqref{correlation} to be rewritten as
\begin{equation}
\label{correlation2}
    \sum_{k=1}^N \hspace{-0.03cm}\hat{\mathbf{\Phi}}(t_k, \hspace{-0.02cm}\boldsymbol{\beta})\mathbf{\Sigma}^{-\hspace{-0.03cm}1}\hspace{-0.03cm}(\bm{\beta})\hspace{-0.03cm}\Bigl(\mathbf{\Upsilon}(\hspace{-0.04cm}t_k,\hspace{-0.02cm} \boldsymbol{\beta})\hspace{-0.04cm}-\hspace{-0.05cm} \mathbf{\Phi}^\top\hspace{-0.04cm}(t_k,\hspace{-0.02cm}\boldsymbol{\beta})\mathcal{B} \Bigr) \hspace{-0.06cm}=\hspace{-0.03cm} \mathbf{0},
\end{equation}
where $\mathcal{B}$ is a block diagonal matrix with diagonal elements given by the submodel parameter vectors in $\boldsymbol{\beta}$; $\mathbf{\Upsilon}(t_k, \boldsymbol{\beta})$ is formed by stacking the filtered variants of the $K$ residual output vectors in \eqref{eq: CT - residual output subsystem} as columns; and $\mathbf{\Phi}(t_k, \boldsymbol{\beta})$ is constructed by stacking the $K$ regressor matrices from \eqref{eq: CT - regressor matrix subsystem} as rows. A solution to \eqref{correlation2} is found iteratively by fixing at the $j$th iteration $\boldsymbol{\beta} = \boldsymbol{\beta}^{\langle j \rangle}$ in the residual output matrix $\mathbf{\Upsilon}(t_k, \boldsymbol{\beta})$, the regressor matrix $\mathbf{\Phi}(t_k,\boldsymbol{\beta})$, and the instrument matrix $\hat{\mathbf{\Phi}}(t_k, \boldsymbol{\beta})$, and solving for $\mathcal{B}$:
\begin{align}
\mathcal{B}^{\langle j+1 \rangle} = &\left[ \sum_{k=1}^N \hat{\mathbf{\Phi}}(t_k, \boldsymbol{\beta}^{\langle j \rangle}) \mathbf{\Sigma}^{-1}(\boldsymbol{\beta}^{\langle j \rangle}) \mathbf{\Phi}^{\top}(t_k, \boldsymbol{\beta}^{\langle j \rangle})\right]^{-1}  \notag \\
\label{eq: CT - estimator}
&\times \sum_{k=1}^N \hat{\mathbf{\Phi}}(t_k, \boldsymbol{\beta}^{\langle j \rangle}) \mathbf{\Sigma}^{-1}(\boldsymbol{\beta}^{\langle j \rangle}) \mathbf{\Upsilon}(t_k, \boldsymbol{\beta}^{\langle j \rangle}),
\end{align}
where we have used the parameter vector of the current iteration to obtain a sample maximum likelihood estimate \cite{Pintelon2012SystemIdentification} of the covariance matrix $\bm{\Sigma}_0$ as
\begin{equation} \label{eq: CT - ML noise covariance}
    \mathbf{\Sigma}(\boldsymbol{\beta}^{\langle j \rangle}) = \frac{1}{N} \sum_{k=1}^N \boldsymbol{\varepsilon}(t_k,\boldsymbol{\beta}^{\langle j \rangle})\boldsymbol{\varepsilon}^\top(t_k,\boldsymbol{\beta}^{\langle j \rangle}).
\end{equation}

The next iteration of parameter vector $\boldsymbol{\beta}^{\langle j+1\rangle}$ is extracted from the block diagonal coefficients of $\mathcal{B}^{\langle j+1\rangle}$, and the converging point of this iterative procedure provides for an estimate of the solution to \eqref{correlation2}.

\vspace{-0.1cm}
\begin{rem}
    The procedure in \eqref{eq: CT - estimator} reduces to the SRIVC method \cite{Young1980RefinedAnalysis} for open-loop systems and the CLSRIVC method \cite{gilson2008instrumental} for closed-loop systems when $K=1$ and the system is SISO. It simplifies to a time-domain variant of the instrumental variables method in \cite{Blom2010MultivariableRegression} when the system is MIMO and non-additive, and it is equivalent to the method in \cite{Gonzalez2024IdentificationClosed-loop} when $K>1$ and the system is SISO.
\end{rem}

\section{Statistical analysis} 
\label{sec:statistical}
Understanding the asymptotic properties of the developed estimator is essential for ensuring reliability and accuracy in stochastic settings. This section analyzes two properties of the estimator: generic consistency \cite{soderstrom1984generic} and the asymptotic distribution of the parameter estimates. 

\subsection{Generic consistency}
In this subsection, we prove that the estimator derived in Section \ref{sec:riv} is generically consistent, meaning it converges to the true parameter vector with probability one for almost all asymptotically stable LTI system and model parameters, except possibly for special cases with Lebesgue measure zero in the parameter space \cite{soderstrom1984generic}. Establishing this result requires the following set of assumptions, stated separately for the open- and closed-loop settings.
\begin{assumption}[Persistency of excitation] \label{assumption1} Open loop: The input $\mathbf{u}(t_k)$ is persistently exciting \cite[p. 415]{Ljung1999SystemUser} of order no less than $2\sum_{i=1}^K n_i$. Closed loop: The reference $\mathbf{r}(t_k)$ is persistently exciting of order no less than $n_{\textnormal{c}}+2\sum_{i=1}^K n_i$, where $n_{\textnormal{c}}$ is the order of the discrete-time controller $\mathbf{C}_{\mathrm{d}}(q)$. \label{assumption: 1 - CT - consistency}
\end{assumption}
\begin{assumption}[Stationarity and independence] \label{assumption2} Open loop: The input signal $\mathbf{u}(t_k)$ is quasi-stationary \cite[p. 34]{Ljung1999SystemUser} and statistically independent of $\mathbf{v}(t_s)$  for all $k$ and $s$. Closed-loop: The reference $\mathbf{r}(t_k)$ is quasi-stationary
and statistically independent of $\mathbf{v}(t_s)$ for all $k$ and $s$.
\end{assumption}

\begin{assumption}[Stability and coprimeness] \label{assumption3} For all $j\geq 1$ and $i=1,\dots,K$, every root $s$ of the model denominator estimate $A^{\langle j \rangle}_{i}(p)$ has a strictly negative real part and satisfies $\mathbf{B}_{i}^{\langle j \rangle}(s)\neq \mathbf{0}$. Closed loop: Additionally, the discrete-time equivalent model estimates at each iteration are stabilized by the controller $\mathbf{C}_{\mathrm{d}}(q)$.
\end{assumption}

\begin{assumption}[Sampling rate]\label{assumption4} For all $j\geq 1$, the sampling frequency is more than twice that of the largest imaginary part of the zeros of $\prod_{i=1}^K A^{\langle j \rangle}_i(p) A_i^*(p)$. \label{assumption: 4 - CT - consistency}
\end{assumption}

Assumptions \ref{assumption1} and \ref{assumption2} are standard in the statistical analysis of algorithms for open and closed-loop identification \cite{Soderstrom1983InstrumentalIdentification}. Assumption \ref{assumption3} can be relaxed to admit the identification of unstable systems if an additional all-pass filter is incorporated at each iteration as in \cite{Gonzalez2023RefinedClosed-loop}, and Assumption \ref{assumption4} is required to ensure a bijective discrete-time to continuous-time transformation \cite{kollar1996equivalence}.

The main theorem of this section involves decomposing the regressor submatrices $\mathbf{\Phi}_{i}\left(t_k, \boldsymbol{\beta}\right)$ in \eqref{eq: CT - regressor matrix subsystem} in three parts. Specifically, we consider
\begin{equation} \label{eq: CT - consistency - decomposition regressor}
\mathbf{\Phi}_{i}\left(t_k, \boldsymbol{\beta}\right) = \tilde{\mathbf{\Phi}}_{i}^{\mathbf{z}}\left(t_k, \boldsymbol{\beta}\right) + \mathbf{\Delta}_i^{\mathbf{z}}\left(t_k, \boldsymbol{\beta}\right) -\mathbf{v}_i^{\mathbf{z}}\left(t_k, \boldsymbol{\beta}\right).
\end{equation}
Here, $\tilde{\mathbf{\Phi}}_{i}^{\mathbf{z}}\left(t_k, \boldsymbol{\beta}\right)$ denotes the noise and interpolation error-free regressor matrix
\begin{align} \label{eq: CT - consistency - error-free regressor}
\tilde{\mathbf{\Phi}}^{\mathbf{z}}_{i}&\left(t_k, \boldsymbol{\beta}\right) = \bigg[ \dfrac{-p\mathbf{B}^*_i(p)}{A_i(p)A^*_i(p)}\mathbf{z}(t_k), \dots,\ \dfrac{-p^{n_i}\mathbf{B}^*_i(p)}{A_i(p)A^*_i(p)}\mathbf{z}(t_k), \notag \\
&\dfrac{1}{A_i(p)}\mathbf{Z}^\top(t_k), \dots,\ \dfrac{p^{m_i}}{A_i(p)}\mathbf{Z}^\top(t_k)  \bigg]^{\top},
\end{align}
where $\mathbf{z}(t_k)$ is equal to $\mathbf{u}(t_k)$ and in the open-loop case, and $\mathbf{S}_{uo}^*(q)\mathbf{r}(t_k)$ in the closed-loop case. The perturbation matrix $\boldsymbol{\Delta}_i^{\mathbf{z}}\left(t_k, \boldsymbol{\beta}\right)$ takes the form
\begin{equation} \label{eq: CT consistency - perturbation matrix}
\boldsymbol{\Delta}_i^{\hspace{-0.02cm}\mathbf{z}}\hspace{-0.02cm}(\hspace{-0.01cm}t_k, \hspace{-0.02cm}\boldsymbol{\beta}) \hspace{-0.09cm}=\hspace{-0.11cm} \left[\hspace{-0.02cm}\boldsymbol{\Delta}_{i, 1}^{\mathbf{z}}\hspace{-0.04cm}(\hspace{-0.01cm}t_k,\hspace{-0.02cm} \boldsymbol{\beta}),\hspace{-0.02cm} \dots,\hspace{-0.02cm}\boldsymbol{\Delta}_{i, n_i}^{\mathbf{z}}\hspace{-0.05cm}(\hspace{-0.01cm}t_k,\hspace{-0.02cm} \boldsymbol{\beta}), \hspace{-0.02cm}\mathbf{0}, \hspace{-0.02cm}\dots, \hspace{-0.02cm}\mathbf{0}\right]^{\hspace{-0.03cm}\top}\hspace{-0.12cm}, \hspace{-0.05cm}
\end{equation}
where the entries for $j = 1,\ \dots,\ n_i$ are given by
\begin{align} 
&\boldsymbol{\Delta}_{i, j}^{\mathbf{z}}\hspace{-0.01cm}(t_k, \hspace{-0.02cm}\boldsymbol{\beta})\hspace{-0.07cm} = \hspace{-0.07cm}\frac{p^j \mathbf{B}_i^*(p)}{A_i\hspace{-0.02cm}(p) A_i^*\hspace{-0.02cm}(p)} \mathbf{z}(t_k) \hspace{-0.06cm}-\hspace{-0.07cm}  \frac{p^j}{A_i(p)}\hspace{-0.07cm}\left\{\hspace{-0.04cm}\frac{\mathbf{B}_i^*\hspace{-0.02cm}(p)}{A_i^*\hspace{-0.02cm}(p)} \mathbf{z}(t)\hspace{-0.04cm}\right\}_{\hspace{-0.04cm}t=t_k}  \notag \\
& \hspace{0.5cm}-\sum_{\substack{\ell=1,\ldots, K \\
\ell \neq i}} \frac{p^j}{A_i(p)}\big\{\hspace{-0.05cm}\left(\mathbf{G}_\ell^*(p)-\mathbf{G}_\ell(p)\right) \mathbf{z}(t)\big\}_{t=t_k},
\end{align}
where the notation $\mathbf{H}(p)\{\mathbf{G}(p)\mathbf{x}(t)\}_{t=t_k}$ means that the filtered signal $\mathbf{G}(p)\mathbf{x}(t)$ (with $\mathbf{x}(t)$ having ZOH intersample behavior) is sampled prior to being filtered by $\mathbf{H}(p)$ as in Remark \ref{rem1}. Finally, the contribution of the noise to the regressor, $\mathbf{v}_i^{\mathbf{z}}(t_k, \boldsymbol{\beta})$, is given by filtered versions of $\mathbf{v}(t_k)$. Their exact expressions are omitted since these are not relevant to the upcoming statistical analysis.

Theorem \ref{theorem: CT - consistency} shows that the true parameter vector of the additive system and the noise covariance are recovered with probability one under mild conditions as the number of samples tends to infinity.

\begin{theorem}[Generic consistency] \label{theorem: CT - consistency}
Consider the proposed estimator (\ref{eq: CT - estimator}) for the open- and closed-loop settings, and suppose Assumptions \ref{assumption: 1 - CT - consistency} to \ref{assumption: 4 - CT - consistency} hold. Then, the following statements are true:
\begin{enumerate}
    \item[1.] The matrix $\overline{\mathbb{E}}\{\hat{\mathbf{\Phi}}\left(t_k, \boldsymbol{\beta}\right) \mathbf{\Sigma}^{-1}(\boldsymbol{\beta})\mathbf{\Phi}^{\top}\left(t_k, \boldsymbol{\beta}\right)\}$ is generically nonsingular with respect to the system and model denominator polynomials, provided that the following condition holds:
    
\begin{align} 
            &\left\|\overline{\mathbb{E}}\left\{\hat{\mathbf{\Phi}}\left(t_k, \boldsymbol{\beta}\right)\mathbf{\Sigma}^{-1}(\boldsymbol{\beta}) \mathbf{\Delta}^{\mathbf{z} \top}\left(t_k, \boldsymbol{\beta}\right)\right\}\right\|_2\notag \\
            \label{eq: consistency - cond 1}&\hspace{0.0cm}<\hspace{-0.03cm}\sigma_{\min }\hspace{-0.06cm}\left(\overline{\mathbb{E}}\hspace{-0.03cm}\left\{\hspace{-0.03cm}\hat{\boldsymbol{\mathbf{\Phi}}}\left(t_k, \boldsymbol{\beta}\right)\mathbf{\Sigma}^{-\hspace{-0.02cm}1}\hspace{-0.03cm}(\boldsymbol{\beta}) \tilde{\mathbf{\Phi}}^{\mathbf{z}  \top}\left(t_k, \boldsymbol{\beta}\right)\right\}\right)\hspace{-0.03cm},
            \end{align}
             where $\hat{\mathbf{\Phi}}(t_k, \boldsymbol{\beta}), \tilde{\mathbf{\Phi}}^{\mathbf{z}}(t_k, \boldsymbol{\beta})$, and $\boldsymbol{\Delta}^{\mathbf{z}}(t_k, \boldsymbol{\beta})$ are formed by stacking the matrices described by \eqref{instrumentmatrix}, \eqref{eq: CT - consistency - error-free regressor} and \eqref{eq: CT consistency - perturbation matrix}, respectively.
    
    \item[2.] Assume that (\ref{eq: consistency - cond 1}) is satisfied, and the iterations of the estimator converge for all $N$ sufficiently large to say, $\hat{\mathcal{B}}_N$, with $\hat{\mathcal{B}}_N$ being a block-diagonal matrix formed by $\hat{\boldsymbol{\beta}}_N$. Then, as $N \to \infty$, the additive model defined by $\hat{\boldsymbol{\beta}}_N$ converges to the true additive system, and $\bm{\Sigma}(\hat{\boldsymbol{\beta}}_N) \to \bm{\Sigma}_0$ with probability~1.
    \end{enumerate}
\end{theorem}
\textit{Proof of Statement 1}: The decomposition in \eqref{eq: CT - consistency - decomposition regressor} leads to
\begin{align}
    \overline{\mathbb{E}}&\left\{\hat{\boldsymbol{\Phi}}(t_k,\boldsymbol{\beta}) \mathbf{\Sigma}^{-1}(\boldsymbol{\beta})\boldsymbol{\Phi}^{\top}(t_k, \boldsymbol{\beta})\right\} \notag \\
    &\hspace{0.6cm}= \overline{\mathbb{E}}\left\{\hat{\boldsymbol{\Phi}}(t_k, \boldsymbol{\beta}) \mathbf{\Sigma}^{-1}(\boldsymbol{\beta}) \tilde{\mathbf{\Phi}}^{\mathbf{z}\top}(t_k, \boldsymbol{\beta})\right\} \notag \\
    \label{eq: consistency - modified normal matrix decomposition}
    &\hspace{0.6cm}+ \overline{\mathbb{E}}\left\{\hat{\boldsymbol{\Phi}}(t_k, \boldsymbol{\beta})\mathbf{\Sigma}^{-1}(\boldsymbol{\beta}) \mathbf{\Delta}^{\mathbf{z}\top}(t_k, \boldsymbol{\beta})\right\},
\end{align}
where the expectations are well-defined under the quasi-stationarity assumption in Assumption \ref{assumption2}. The noise-dependent term in the decomposition vanishes, since the instrument  $\hat{\boldsymbol{\Phi}}\left(t_k, \boldsymbol{\beta}\right)$ is uncorrelated with the noise term $ \mathbf{v}^{\mathbf{z}}\left(t_k, \boldsymbol{\beta}\right)$ for fixed $\bm{\beta}$. Assuming that the condition in \eqref{eq: consistency - cond 1} holds, Theorem 5.1 of \cite{dahleh2002lectures} implies that the perturbation matrix $\overline{\mathbb{E}}\{\hat{\boldsymbol{\Phi}}\left(t_k, \boldsymbol{\beta}\right)\mathbf{\Sigma}^{-1}(\boldsymbol{\beta})  \mathbf{\Delta}^{\mathbf{z}\top}\left(t_k, \boldsymbol{\beta}\right)\}$ is sufficiently small in 2-norm such that it will not affect the generic nonsingularity of the matrix $\overline{\mathbb{E}}\{\hat{\boldsymbol{\Phi}}\left(t_k, \boldsymbol{\beta}\right) \mathbf{\Sigma}^{-1}(\boldsymbol{\beta}) \mathbf{\Phi}^\top\left(t_k, \boldsymbol{\beta}\right)\}$. Therefore, it suffices to prove the generic nonsingularity of the matrix $\overline{\mathbb{E}}\{\hat{\boldsymbol{\Phi}}\left(t_k, \boldsymbol{\beta}\right) \mathbf{\Sigma}^{-1}(\boldsymbol{\beta}) \tilde{\mathbf{\Phi}}^{\mathbf{z}\top}\left(t_k, \boldsymbol{\beta}\right)\}$.

The noise-free, interpolation-error-free regressor submatrices (\ref{eq: CT - consistency - error-free regressor}) are written as
\begin{equation} \label{eq: CT - consistency - decomposition phi}
\tilde{\mathbf{\Phi}}_{i}^{\mathbf{z}}\left(t_k, \boldsymbol{\beta}\right) = \frac{1}{A_{i}(p) A_i^*(p)} \mathbf{R}_i(t_k, \boldsymbol{\beta}),
\end{equation}
with 
\begin{align} \label{eq: CT - consistency - R matrix}
\mathbf{R}_i(t_k, \boldsymbol{\beta})  &=\ \Bigl[ -p\mathbf{B}_i^*(p) \mathbf{z}(t_k), \dots, -p^{n_i}\mathbf{B}_i^*(p) \mathbf{z}(t_k), \notag \\
& A_i^*(p)\mathbf{Z}^\top(t_k), \dots, p^{m_i}A_i^*(p)\mathbf{Z}^\top(t_k) \Bigr]^{\top}.
\end{align}
Each entry of $\mathbf{R}_i(t_k, \boldsymbol{\beta})$ associated with the numerator polynomial can be rewritten as $\mathbf{B}^*_{i}(p) \mathbf{z}(t_k) = \mathbf{Z}^{\top}(t_k)\operatorname{vec}\{\mathbf{B}^*_{i}(p) \}$, which enables $\mathbf{R}_i(t_k, \boldsymbol{\beta})$ to be expressed as the product between a rectangular Sylvester matrix and a matrix containing the derivatives of the input signals, i.e.,
\begin{equation}
\label{RiwithZ}
\mathbf{R}_i\left(t_k, \boldsymbol{\beta}\right) = \mathbf{S}(-\mathbf{B}_{i}^*, A_i^*) \bm{\mathcal{Z}}_{i}(t_k),
\end{equation}
where the rectangular Sylvester matrix is defined as

\begin{widetext}
\begin{equation} \label{eq: CT - consistency - sylvester matrix}
\mathbf{S}(-\mathbf{B}_i^*, A_i^*) = \left[\begin{array}{cccccc}
\mathbf{0} & & -\operatorname{vec}\left\{\mathbf{B}^*_{i,m_i} \right\}^{\top}  & \ldots & -\operatorname{vec}\left\{\mathbf{B}^*_{i,0} \right\}^{\top} & \mathbf{0} \\
& \text{\reflectbox{$\ddots$}} & & \text{\reflectbox{$\ddots$}} & \text{\reflectbox{$\ddots$}} & \\
-\operatorname{vec}\left\{\mathbf{B}^*_{i,m_i} \right\}^{\top}  & \ldots & -\operatorname{vec}\left\{\mathbf{B}^*_{i,0} \right\}^{\top} & \mathbf{0} & & \mathbf{0} \\
\hline \mathbf{0} & & a_{n_i}^*\mathbf{I}_{n_\textnormal{u}n_\textnormal{y}}  & \ldots & a_1^*\mathbf{I}_{n_\textnormal{u}n_\textnormal{y}} & \mathbf{I}_{n_\textnormal{u}n_\textnormal{y}} \\
& \text{\reflectbox{$\ddots$}}  & & \text{\reflectbox{$\ddots$}} & \text{\reflectbox{$\ddots$}} & \\
a_{n_i}^*\mathbf{I}_{n_\textnormal{u}n_\textnormal{y}} & \ldots & a_1^*\mathbf{I}_{n_\textnormal{u}n_\textnormal{y}} & \mathbf{I}_{n_\textnormal{u}n_\textnormal{y}} & & \mathbf{0}
\end{array}\right],
\end{equation}
\end{widetext}
\hspace{-0.27cm}and the $n_\textnormal{u}n_\textnormal{y}(n_i+m_i+1) \times n_\textnormal{y}$ dimensional matrix $\bm{\mathcal{Z}}_{i}\left(t_k\right) = \mathbf{M}_i(p)\mathbf{Z}(t_k)$, where 
\begin{equation} \label{M}
\mathbf{M}_i(p) \hspace{-0.07cm}=\hspace{-0.07cm} \left[p^{n_i+m_i}, p^{n_i+m_i-1},\dots,1\right]^{\top}\otimes \mathbf{I}_{n_\textnormal{u}n_\textnormal{y}}. 
\end{equation}
For the instrument matrix the same procedure is completed, which results in the form
\begin{align}
\hat{\boldsymbol{\Phi}}_{i}\left(t_k, \boldsymbol{\beta}\right) = \mathbf{S}(-\mathbf{B}_i,A_i) \frac{1}{A_i^2(p)} \bm{\mathcal{Z}}_{i}(t_k).
\end{align}
The Sylvester matrices $\mathbf{S}(-\mathbf{B}_i^*, A_i^*)$ and $\mathbf{S}(-\mathbf{B}_i, A_i)$ are permutations of the Sylvester matrix structure presented in \cite[Eq. A3.9]{Soderstrom1983InstrumentalIdentification}. Consequently, under the coprimeness condition stated in Assumption \ref{assumption3}, they are guaranteed to have full row rank according to Lemma A3.2 of \cite{Soderstrom1983InstrumentalIdentification}. Next, the block diagonal matrices $\bm{\mathcal{S}}^*$ and $\bm{\mathcal{S}}$ are introduced, which are block diagonal matrices formed by $\{\mathbf{S}(-\mathbf{B}_i^*, A_i^*)\}_{i=1}^K$ and $\{\mathbf{S}(-\mathbf{B}_i, A_i)\}_{i=1}^K$ on their respective diagonals. Using these matrices, we find
\begin{equation} \label{eq: CT - consistency - Sylvester decomposition}
 \overline{\mathbb{E}}\left\{\hat{\boldsymbol{\Phi}}(t_k, \boldsymbol{\beta})\mathbf{\Sigma}^{-1}(\boldsymbol{\beta})  \tilde{\mathbf{\Phi}}^{\mathbf{z}\top}(t_k, \boldsymbol{\beta})\right\} = \bm{\mathcal{S}} \mathbf{\Psi}_\mathbf{z} {\bm{\mathcal{S}}^*}^{\top},
\end{equation}
where
\begin{equation} \label{eq: CT - consistency - psy}
\boldsymbol{\Psi}_{\hspace{-0.04cm}\mathbf{z}}\hspace{-0.11cm}=\hspace{-0.08cm}\overline{\mathbb{E}}\hspace{-0.07cm}\left\{\hspace{-0.11cm}\left[\begin{array}{c}
\hspace{-0.1cm}\dfrac{1}{{A^{2}_{1}}(p)} \bm{\mathcal{Z}}_1(t_k) \\
\vdots\\
\hspace{-0.11cm}\dfrac{1}{{A_{\hspace{-0.03cm}K}^2}\hspace{-0.03cm}(p)} \bm{\mathcal{Z}}_{\hspace{-0.035cm}K}\hspace{-0.025cm}(\hspace{-0.01cm}t_k\hspace{-0.01cm})
\end{array}\hspace{-0.1cm}\right]\hspace{-0.13cm}\mathbf{\Sigma}^{\hspace{-0.02cm}-\hspace{-0.03cm}1}\hspace{-0.04cm}(\boldsymbol{\beta})\hspace{-0.15cm}\left[\hspace{-0.11cm}\begin{array}{c}
\dfrac{1}{A_1^*(p) \hspace{-0.02cm}A_1\hspace{-0.02cm}(p)} \bm{\mathcal{Z}}^\top_1\hspace{-0.03cm}(t_k) \\
\vdots \\
\dfrac{1}{A_{\hspace{-0.04cm}K}^*\hspace{-0.02cm}(p) \hspace{-0.02cm}A_{\hspace{-0.03cm}K}\hspace{-0.03cm}(\hspace{-0.01cm}p\hspace{-0.01cm})} \bm{\mathcal{Z}}^{\top}_{\hspace{-0.04cm}K}\hspace{-0.02cm}(\hspace{-0.005cm}t_k\hspace{-0.005cm})
\end{array}\hspace{-0.1cm}\right]^{\hspace{-0.12cm}\top}\hspace{-0.07cm}\right\}\hspace{-0.05cm}.
\end{equation}
The real analyticity of the entries of $\bm{\mathcal{S}} \mathbf{\Psi}_\mathbf{z} {\bm{\mathcal{S}}^*}^{\top}$ can be established using the same argument as in Lemma 9 of \cite{Pan2020ConsistencySystems}. Thus, the generic nonsingularity of $\bm{\mathcal{S}} \mathbf{\Psi}_\mathbf{z} {\bm{\mathcal{S}}^*}^{\top}$ follows by showing that such matrix is nonsingular when evaluated at the true system parameters $\boldsymbol{\beta}^*$ \cite[Lemma A2.3]{Soderstrom1983InstrumentalIdentification}. At $\bm{\beta}=\bm{\beta}^*$, \eqref{eq: CT - consistency - Sylvester decomposition} reduces to the positive semidefinite matrix $\bm{\mathcal{S}}^*\mathbf{\Psi}_\mathbf{z}^* {\bm{\mathcal{S}}^*}^{\top}$, which is nonsingular if and only if $\mathbf{\Psi}_\mathbf{z}$ is nonsingular since $\bm{\mathcal{S}}^*$ has full row rank. From Lemma \ref{lemma: pos def} in the Appendix, $\mathbf{\Psi}_\mathbf{z}$ is positive definite when $\boldsymbol{\beta} = \boldsymbol{\beta}^*$, and thus nonsingular. Consequently, $\bm{\mathcal{S}}\mathbf{\Psi}_\mathbf{z} {\bm{\mathcal{S}}^*}^{\top}$ is generically nonsingular, which completes the proof. \hfill $\square$

\textit{Proof of Statement 2}: Let $\bar{\mathcal{B}}$ be the limiting point of $\hat{\mathcal{B}}_N$ as $N$ goes to infinity, with block diagonal elements being described by $\bar{\bm{\beta}}$. By letting $N \rightarrow \infty$ in \eqref{eq: CT - estimator}, the ergodicity theorems in \cite[Appendix 2]{Ljung1999SystemUser} lead to the condition
\begin{align} 
\overline{\mathbb{E}}&\left\{ \hat{\mathbf{\Phi}}(t_k, \bar{\boldsymbol{\beta}}) \bm{\Sigma}^{-1}(\bar{\bm{\beta}}) \mathbf{\Phi}^{\top}(t_k, \bar{\boldsymbol{\beta}})\right\}^{-1}  \notag \\
\label{secondterm}
&\times \overline{\mathbb{E}}\left\{ \hat{\mathbf{\Phi}}(t_k, \bar{\boldsymbol{\beta}})  \bm{\Sigma}^{-1}(\bar{\bm{\beta}})  \boldsymbol{\varepsilon}(t_k, \bar{\boldsymbol{\beta}}) \right\} = \mathbf{0}, 
\end{align}
where we have used the fact that $\boldsymbol{\Upsilon}(t_k, \bar{\boldsymbol{\beta}})-\mathbf{\Phi}^{\top}(t_k, \bar{\boldsymbol{\beta}})\bar{\mathcal{B}}=\boldsymbol{\varepsilon}(t_k,\bar{\boldsymbol{\beta}})[1,\dots,1]$. Since the matrix being inverted in \eqref{secondterm} is generically nonsingular by Statement 1, the limiting point $\bar{\bm{\beta}}$ leads to the second term in \eqref{secondterm} being identically zero. By introducing the matrix polynomial $\mathbf{H}_i(p) = A_i^*(p) \bar{\mathbf{B}}_i(p)-\bar{A}_i(p)\mathbf{B}_i^*(p)$, the residual function is rewritten according to
\begin{align}
    \boldsymbol{\varepsilon}(t_k,\bar{\boldsymbol{\beta}})&= \sum_{i=1}^K\Bigl(\mathbf{G}^*_i(p)-\mathbf{G}_i(p, \bar{\boldsymbol{\theta}}_i)\Bigr) \mathbf{u}(t_k) + \mathbf{v}(t_k), \notag \\
     &= \sum_{i=1}^K \frac{\mathbf{H}_i(p)}{A_i^*(p) \bar{A}_i(p)} \mathbf{z}(t_k) + \tilde{\mathbf{v}}(t_k), \label{eq: CT - consistency - rewritten residual}
\end{align}
where the noise contribution, defined as $\tilde{\mathbf{v}}(t_k)$ in \eqref{eq: CT - consistency - rewritten residual}, is uncorrelated with $\mathbf{z}(t_k)$. Together with the fact that the inverse covariance matrix $\bm{\Sigma}^{-1}(\overline{\bm{\beta}})$ converges with probability 1 to a deterministic matrix as $N\to\infty$ \cite[Theorem 2.3]{Ljung1999SystemUser}, we obtain $\overline{\mathbb{E}}\{\hat{\mathbf{\Phi}}(t_k, \bar{\boldsymbol{\beta}}) \bm{\Sigma}^{-1}(\bar{\bm{\beta}}) \tilde{\mathbf{v}}(t_k)\}=\mathbf{0}$. By following the same steps leading to \eqref{RiwithZ}, we find that $\mathbf{H}_i(p)\mathbf{z}(t_k)=\bm{\mathcal{Z}}_i^\top(t_k)\boldsymbol{\eta}_i$, where $\boldsymbol{\eta}_i$ contains the coefficients of $\mathbf{H}_i(p)$ in descending order of degree
\begin{equation}
    \boldsymbol{\eta}_i \hspace{-0.1cm}=\hspace{-0.14cm} \Bigl[\hspace{-0.01cm}\operatorname{vec}\hspace{-0.01cm}\{\hspace{-0.01cm}\mathbf{H}_{i,n_i\hspace{-0.02cm}+\hspace{-0.02cm}m_i}\hspace{-0.01cm}\}^{\hspace{-0.05cm}\top}\hspace{-0.06cm}, \hspace{-0.02cm}\operatorname{vec}\{\mathbf{H}_{i,n_i\hspace{-0.02cm}+\hspace{-0.02cm}m_i\hspace{-0.02cm}-\hspace{-0.02cm}1}\hspace{-0.01cm}\}^{\hspace{-0.05cm}\top}\hspace{-0.04cm}, \hspace{-0.02cm}\dots\hspace{-0.02cm},\hspace{-0.02cm} \operatorname{vec}\{\hspace{-0.01cm}\mathbf{H}_{i,\hspace{-0.02cm}0}\}^{\hspace{-0.04cm}\top} \hspace{-0.02cm}\Bigr]^{\hspace{-0.06cm}\top}\hspace{-0.1cm}. \notag 
\end{equation}
If $\boldsymbol{\eta}$ is the vector obtained by stacking $\boldsymbol{\eta}_i$ for $i = 1,\dots,K$, then we can write
\begin{equation} \label{eq: CT - consistency - last term}
\overline{\mathbb{E}}\left\{ \hat{\mathbf{\Phi}}(t_k, \bar{\boldsymbol{\beta}}) \bm{\Sigma}^{-1}(\bar{\bm{\beta}})\boldsymbol{\varepsilon}(t_k, \bar{\boldsymbol{\beta}}) \right\} = \bar{\bm{\mathcal{S}}} \bar{\mathbf{\Psi}}_\mathbf{z}\boldsymbol{\eta}, 
\end{equation}
where $\bar{\bm{\mathcal{S}}}$ is a block diagonal matrix formed by the Sylvester matrices $\mathbf{S}(-\bar{\mathbf{B}}_i, \bar{A}_i)$ and the matrix $\bar{\mathbf{\Psi}}_\mathbf{z}$ is given by \eqref{eq: CT - consistency - psy} when evaluated at $\bm{\beta}=\bar{\boldsymbol{\beta}}$. The matrix $\bar{\bm{\mathcal{S}}}$ has full column rank due to the coprimeness of $\bar{\mathbf{B}}_i$ and $\bar{A}_i$ by Assumption \ref{assumption3}, and by Statement 1, $\bar{\mathbf{\Psi}}_\mathbf{z}$ is generically nonsingular. Thus, for \eqref{eq: CT - consistency - last term} to become identically zero, it must hold that $\boldsymbol{\eta} = \mathbf{0}$, which directly implies that $\mathbf{G}_i(p, \bar{\boldsymbol{\theta}}_i) = \mathbf{G}_i^*(p)$ for $i = 1,...,K$. To conclude the proof, Lemma \ref{lemmaqss} in Appendix \ref{appendixA} leads to 
$\bm{\Sigma}(\hat{\bm{\beta}}_N)\xrightarrow{} \bm{\Sigma}_0$ with probability 1 as $N\to\infty$, generically on the system and model denominator parameters. \hfill $\square$

\begin{rem}
The statistical analysis does not cover the case of marginally stable systems, as the potential loss of quasi-stationarity in the signals complicates the convergence analysis of sample covariances. While the methods and analysis could be extended to this scenario, as well as to unstable systems \cite{Gonzalez2023RefinedClosed-loop}, such extensions are beyond the scope of this paper.
\end{rem}

\subsection{Asymptotic distribution}

In this subsection, we derive an expression for the asymptotic covariance of the estimator resulting from the iterations in \eqref{eq: CT - estimator}. Importantly, as mentioned in Section \ref{sec:riv}, the underlying additive system is identifiable only up to block permutations of the parameter vector, meaning the true system corresponds to an equivalence class of parameter vectors. Let $\boldsymbol{\beta}^*$ denote a fixed representative of this class, with structure given in \eqref{eq: CT - parameter vector}.

To make the asymptotic analysis well-defined, we assume that the estimated submodels are ordered such that the resulting estimated parameter vector $\hat{\boldsymbol{\beta}}_N$ is closest to $\boldsymbol{\beta}^*$ in Euclidean norm. When this occurs, we say that the ordering of estimated submodels \emph{aligns} with $\boldsymbol{\beta}^*$.

\begin{theorem}[Asymptotic distribution] 
\label{thmefficiency}
Consider the system output in \eqref{eq: CT - output observation}, where the noise $\mathbf{v}(t_k)$ is independent and identically distributed Gaussian white noise with covariance matrix $\bm{\Sigma}_0$. Assume that the iterations in \eqref{eq: CT - estimator} converge for all $N$ sufficiently large, resulting in the estimator $\hat{\bm{\beta}}_N$, and that the ordering of submodels in $\hat{\bm{\beta}}_N$ aligns with $\boldsymbol{\beta}^*$. Then, if Assumptions \ref{assumption: 1 - CT - consistency} to \ref{assumption: 4 - CT - consistency} hold, the estimator $\hat{\boldsymbol{\beta}}_N$ is asymptotically Gaussian distributed, i.e.,
\begin{equation}
\label{asymptoticdistribution}
\sqrt{N}(\hat{\boldsymbol{\beta}}_N - \boldsymbol{\beta}^*) \xrightarrow{\text {dist.}} \mathcal{N}(\mathbf{0}, \mathbf{P}),    
\end{equation}
where
\begin{equation}
\label{pmatrix}
    \mathbf{P}= \overline{\mathbb{E}}\left\{\hat{\mathbf{\Phi}}(t_k,\boldsymbol{\beta}^*)\boldsymbol{\Sigma}_0^{-1} \hat{\mathbf{\Phi}}^\top(t_k,\boldsymbol{\beta}^*)\right\}^{-1}.
\end{equation}
\end{theorem}
\textit{Proof.} The iterations in \eqref{eq: CT - estimator} define a sequence of estimators in $j$ that converges to a solution of \eqref{correlation} for $N$ large enough. Furthermore, this estimator converges to the true parameter vector $\bm{\beta}^*$ due to the alignment assumption and the consistency result in Theorem \ref{theorem: CT - consistency}. By computing the first-order Taylor expansion of the correlation equation in \eqref{correlation} around $\boldsymbol{\beta}^*$, the estimator $\hat{\bm{\beta}}_N$ satisfies
\begin{align}
&\sum_{k=1}^N \frac{\partial}{\partial \bm{\beta}^\top}\big[\hat{\boldsymbol{\Phi}}(t_k,\boldsymbol{\beta}) \boldsymbol{\Sigma}^{-1}(\bm{\beta})\bm{\varepsilon}(t_k, \boldsymbol{\beta})\big]\Big|_{\bm{\beta}=\bm{\beta}^*}(\hat{\boldsymbol{\beta}}_N-\boldsymbol{\beta}^*) \notag \\
\label{eq: CT - Taylor expansion}
&+\sum_{k=1}^N \hat{\boldsymbol{\Phi}}(t_k,\boldsymbol{\beta}^*) \boldsymbol{\Sigma}^{-1}\hspace{-0.04cm}(\bm{\beta}^*) \mathbf{v}(t_k) \hspace{-0.07cm}=\hspace{-0.07cm}o_p(\hspace{-0.01cm}N\|\hat{\boldsymbol{\beta}}_N\hspace{-0.07cm}-\hspace{-0.05cm}\boldsymbol{\beta}^*\|).
\end{align}
The partial derivative in \eqref{eq: CT - Taylor expansion} can be decomposed as 
\begin{align}
    \frac{\partial}{\partial \bm{\beta}^\top}&\big[\hat{\boldsymbol{\Phi}}(t_k,\boldsymbol{\beta}) \boldsymbol{\Sigma}^{-1}(\bm{\beta})\bm{\varepsilon}(t_k, \boldsymbol{\beta})\big]\Big|_{\bm{\beta}=\bm{\beta}^*} \notag \\
    &= \hspace{-0.03cm}\mathbf{M}(t_k)\mathbf{v}(t_k)\hspace{-0.04cm}-\hspace{-0.05cm}\hat{\boldsymbol{\Phi}}(t_k,\hspace{-0.02cm}\boldsymbol{\beta}^*) \boldsymbol{\Sigma}^{-\hspace{-0.03cm}1}\hspace{-0.04cm}(\bm{\beta}^*)\mathbf{J}(t_k,\boldsymbol{\beta}^*), \notag
\end{align}
where we have used the fact that $\bm{\varepsilon}(t_k, \boldsymbol{\beta}^*)=\mathbf{v}(t_k)$, and where $\mathbf{J}(t_k,\boldsymbol{\beta}^*)$ is the Jacobian of the model output with respect to $\bm{\beta}$, evaluated at the true parameters. The term $\mathbf{M}(t_k)=\frac{\partial}{\partial \bm{\beta}^{\top}}[\hat{\boldsymbol{\Phi}}(t_k,\boldsymbol{\beta}) \boldsymbol{\Sigma}^{-1}(\bm{\beta})]\big|_{\bm{\beta}=\bm{\beta}^*}$ is a tensor of rank 3, and $\mathbf{M}(t_k)\mathbf{v}(t_k)$ is a matrix obtained by tensor-vector contraction. The fact $\sum_{k=1}^N \mathbf{M}(t_k)v(t_k)=o_p(N)$ is shown in Lemma \ref{lemmaMv} of the Appendix, and using the same tools, we find
\begin{equation}
\sum_{k=1}^N \hat{\boldsymbol{\Phi}}(t_k,\boldsymbol{\beta}^*) \left(\boldsymbol{\Sigma}_0^{-1}-\boldsymbol{\Sigma}^{-1}(\bm{\beta}^*)\right)\mathbf{J}(t_k,\boldsymbol{\beta}^*) = o_p(N). \notag
\end{equation}
Thus, \eqref{eq: CT - Taylor expansion} is equivalent to 
\begin{align}
-\frac{1}{N}\sum_{k=1}^N& \hat{\boldsymbol{\Phi}}(t_k,\boldsymbol{\beta}^*) \boldsymbol{\Sigma}_0^{-1}\mathbf{J}(t_k,\boldsymbol{\beta}^*)\sqrt{N}(\hat{\boldsymbol{\beta}}_N-\boldsymbol{\beta}^*) \notag \\
+\frac{1}{\sqrt{N}}&\sum_{k=1}^N \hat{\boldsymbol{\Phi}}(t_k,\boldsymbol{\beta}^*) \boldsymbol{\Sigma}_0^{-1} \mathbf{v}(t_k) =o_p(\sqrt{N}\|\hat{\boldsymbol{\beta}}_N-\boldsymbol{\beta}^*\|). \notag
\end{align}
The first sum converges to its expected value with probability 1 as $N$ tends to infinity due to the quasi-stationarity assumption in Assumption \ref{assumption2}:
\begin{align}
        &\frac{1}{N}\sum_{k=1}^N \hat{\boldsymbol{\Phi}}(t_k,\boldsymbol{\beta}^*) \boldsymbol{\Sigma}_0^{-1}\mathbf{J}(t_k,\boldsymbol{\beta}^*)\notag \\
        &\xrightarrow{\text{w.p.1}} \mathbf{L}:=\overline{\mathbb{E}}\left\{\hat{\boldsymbol{\Phi}}(t_k,\boldsymbol{\beta}^*) \boldsymbol{\Sigma}_0^{-1}\mathbf{J}(t_k,\boldsymbol{\beta}^*)\right\} \textnormal{ as }N\to \infty. \notag
\end{align}
Thus,
\begin{align}
-\mathbf{L}&\sqrt{N}(\hat{\boldsymbol{\beta}}_N-\boldsymbol{\beta}^*)+\frac{1}{\sqrt{N}}\sum_{k=1}^N \hat{\boldsymbol{\Phi}}(t_k,\boldsymbol{\beta}^*) \boldsymbol{\Sigma}_0^{-1} \mathbf{v}(t_k) \notag \\
&=o_p(\sqrt{N}\|\hat{\boldsymbol{\beta}}_N-\boldsymbol{\beta}^*\|). \notag
\end{align}
The matrix $\mathbf{L}$ can be shown to be generically nonsingular by following the same reasoning as in Theorem \ref{theorem: CT - consistency}. Exploiting this fact, standard computations (see, e.g., Theorem 5.21 of \cite{van2000asymptotic}) lead to $\sqrt{N}\|\hat{\boldsymbol{\beta}}_N-\boldsymbol{\beta}^*\| = O_p(1)$, i.e., $\sqrt{N}(\hat{\boldsymbol{\beta}}_N-\boldsymbol{\beta}^*)$ is bounded in probability. Hence, $o_p(\sqrt{N}\|\hat{\boldsymbol{\beta}}_N-\boldsymbol{\beta}^*\|)=o_p(O_p(1))=o_p(1)$, and thus the variant of the central limit theorem in Lemma A4.1 of \cite{Soderstrom1983InstrumentalIdentification} is applicable after premultiplying the last equality by $\mathbf{L}^{-1}$. That is, $\sqrt{N}(\hat{\boldsymbol{\beta}}_N-\boldsymbol{\beta}^*)$ converges in distribution as $N\to \infty$ to a Gaussian distribution
\begin{equation}
\sqrt{N}(\hat{\boldsymbol{\beta}}_N-\boldsymbol{\beta}^*)\xrightarrow{\text { dist. }} \mathcal{N}(\mathbf{0}, \mathbf{L}^{-1}\mathbf{R}\mathbf{L}^{-\top}),
\end{equation}
with $\mathbf{R}$ being given by
\begin{equation} \label{eq: asym. dist. P}
\begin{aligned}
\mathbf{R} & =\lim _{N \rightarrow \infty} \frac{1}{N} \sum_{k=1}^N \sum_{s=1}^N \mathbb{E}\bigg\{\hat{\mathbf{\Phi}}(t_k,\bm{\beta}^*)\boldsymbol{\Sigma}_0^{-1} \mathbf{v}(t_k)\notag \\
&\hspace{1cm}\times\left[\hat{\mathbf{\Phi}}(t_s,\bm{\beta}^*)\boldsymbol{\Sigma}_0^{-1} \mathbf{v}(t_s)\right]^{\top}\bigg\}, \\
& =\lim_{N \rightarrow \infty} \frac{1}{N} \sum_{k=1}^N \sum_{s=1}^N \mathbb{E}\bigg\{\hat{\mathbf{\Phi}}(t_k,\bm{\beta}^*) \boldsymbol{\Sigma}_0^{-1}\mathbb{E}\big\{\mathbf{v}(t_k) \mathbf{v}^\top(t_s)\big\}\notag \\
&\hspace{1cm} \times \boldsymbol{\Sigma}_0^{-1} \hat{\mathbf{\Phi}}^\top(t_s,\bm{\beta}^*)\bigg\}, \\
& = \overline{\mathbb{E}}\left\{\hat{\mathbf{\Phi}}(t_k,\boldsymbol{\beta}^*)\boldsymbol{\Sigma}_0^{-1} \hat{\mathbf{\Phi}}^\top(t_k,\boldsymbol{\beta}^*)\right\}.
\end{aligned}
\end{equation}
For systems operating in open loop, the instrument matrix in \eqref{instrumentmatrix} coincides with the transpose of the Jacobian of the model output, leading to $\mathbf{R}=\mathbf{L}$. For systems operating in closed loop, the Jacobian of the residual can be decomposed in terms of the closed-loop instrument matrix in \eqref{instrumentmatrix} and a term that depends on $\mathbf{S}_{uo}^*(q)\mathbf{v}(t_k)$. This latter term does not contribute to $\mathbf{L}$, since the reference and noise signals are statistically independent according to Assumption \ref{assumption2}. Consequently, $\mathbf{R}=\mathbf{L}$ also for the closed-loop case. Hence, for both scenarios, the asymptotic covariance of $\hat{\bm{\beta}}_N$ is given by $\mathbf{R}^{-1}$, i.e., $\mathbf{P}$ in \eqref{pmatrix}. This concludes the proof. \hfill $\square$

In summary, we have proven that, for either open- or closed-loop setting, the proposed estimator for additive continuous-time models is generically consistent, and is asymptotically normally distributed. The estimator does not require specifying a correct dynamical model for the noise contribution for consistency to hold, which is particularly beneficial for closed-loop applications.

It is important to analyze how the asymptotic covariance in \eqref{asymptoticdistribution} compares to minimum asymptotic covariance among all consistent estimators, namely, the Cram\'er-Rao lower bound (CRLB). Under the assumption that the noise model is correctly assumed to be white and Gaussian, the prediction error method for open loop and the direct approach for closed loop are known to achieve the CRLB at the limit as $N$ tends to infinity \cite{Pan2020EfficiencySystems,forssell1999closed}. For either open- or closed-loop operation, the inverse of the CRLB, also known as the Fisher information matrix \cite{lehmann1998theory}, is given by
\begin{equation}
\label{fisherinformation}
    \mathbf{I}_{\text{F}} =  \overline{\mathbb{E}}\left\{\frac{\partial \boldsymbol{\varepsilon}^\top (t_k,\boldsymbol{\beta})}{\partial \bm{\beta}}\bigg|_{\bm{\beta}=\bm{\beta}^*}\boldsymbol{\Sigma}_0^{-1} \frac{\partial \boldsymbol{\varepsilon}(t_k,\boldsymbol{\beta})}{\partial \bm{\beta}^\top }\bigg|_{\bm{\beta}=\bm{\beta}^*}\right\},
\end{equation}
see e.g., \cite{ljung1993information,forssell1998efficiency} for more details on the discrete-time and SISO system case. This matrix corresponds exactly to \eqref{pmatrix} for the open-loop scenario, which proves that the proposed estimator is asymptotically efficient in open-loop operation. For the closed-loop case, inserting \eqref{uclosedloop} in \eqref{eq: CT - residual function} leads to the matrix identity
\begin{equation}
    \label{matrixinequality}
    \mathbf{I}_{\text{F}} \hspace{-0.05cm}= \hspace{-0.04cm}\mathbf{P}^{-\hspace{-0.02cm}1} \hspace{-0.05cm}+ \overline{\mathbb{E}}\hspace{-0.03cm}\left\{\hspace{-0.03cm}\hat{\mathbf{\Phi}}_{\textnormal{v}}(t_k,\boldsymbol{\beta}^*)\boldsymbol{\Sigma}_0^{-1} \hat{\mathbf{\Phi}}_{\textnormal{v}}^\top(t_k,\boldsymbol{\beta}^*)\right\} \hspace{-0.04cm} \succeq \hspace{-0.03cm}\mathbf{P}^{-1},
    \end{equation}
with $\hat{\mathbf{\Phi}}_{\textnormal{v}}(t_k,\boldsymbol{\beta}^*)$ being the noise component of the closed-loop instrument matrix. Since the expectation in \eqref{matrixinequality} is positive definite under mild conditions, the refined instrumental method in \eqref{correlation} does not achieve the CRLB for the closed-loop case in general. However, by extending the analysis in, e.g., \cite{Soderstrom1983InstrumentalIdentification,soderstrom1987instrumental} to the additive continuous-time MIMO case (see also Theorem 1 of \cite{Gonzalez2024IdentificationClosed-loop}), we find that the estimator reaches the lower bound on the asymptotic covariance of any instrumental variable method \cite{soderstrom1987instrumental}. 

\begin{rem}
\label{remarkcovariance}
In agreement with  Theorem \ref{thmefficiency}, the true covariance matrix of the proposed estimator depends on the true parameter vector and thus in practice it must be estimated using the converging point of the iterative procedure in \eqref{eq: CT - estimator}. Due to the consistency property in Theorem \ref{theorem: CT - consistency}, a consistent estimator for the asymptotic covariance matrix can be formulated as
\begin{equation} \label{eq: CT - estimate asymptotic covariance matrix estimator}
N\textnormal{Cov}(\hat{\bm{\beta}}_N)\hspace{-0.04cm}\approx \hspace{-0.1cm}\left[\hspace{-0.02cm} \frac{1}{N}\hspace{-0.04cm}\sum_{k=1}^N \hspace{-0.03cm}\hat{\boldsymbol{\Phi}}(t_k, \hat{\boldsymbol{\beta}}_N) \boldsymbol{\Sigma}^{-1}\hspace{-0.04cm}(\hat{\boldsymbol{\beta}}_N) \hat{\boldsymbol{\Phi}}^{\top}\hspace{-0.04cm}(t_k, \hat{\boldsymbol{\beta}}_N)\right]^{\hspace{-0.05cm}-\hspace{-0.03cm}1}\hspace{-0.07cm}.
\end{equation}    
\end{rem}

\begin{rem}
    The theoretical guarantees determined in Theorems \ref{theorem: CT - consistency} and \ref{thmefficiency} are general cases of the statistical analyses performed in \cite{Pan2020ConsistencySystems} and \cite{Pan2020EfficiencySystems} for the consistency and asymptotic efficiency SRIVC method, in \cite{Gonzalez2024ConsistencyClosed-loop} for the consistency of the CLSRIVC method, and in \cite{Gonzalez2024IdentificationClosed-loop} for the consistency of the refined instrumental variable method for additive SISO systems in open and closed loop. That is, the results of all these works can be recovered from Theorems \ref{theorem: CT - consistency} and \ref{thmefficiency} of this paper.
\end{rem}

\section{Estimation of structured additive models}
\label{sec:ipem}
The refined instrumental variable method for additive MIMO systems in Section \ref{sec:riv} has been shown to have asymptotically optimal statistical properties for additive model structures with fully parameterized numerator matrix coefficients. However, in many applications, the numerator matrix polynomials are constrained to a lower-dimensional subspace, in which case the method in Section \ref{sec:riv} no longer yields minimum asymptotic covariance. To address this, we first estimate an unstructured additive MIMO model using the approach in Section \ref{sec:riv}, and then project the resulting parameter vector onto the desired subspace of structured models.

The rationale behind the developed method is as follows. Consider the nested model structures $\mathcal{A}$ and $\mathcal{S}$, with $\mathcal{A}$ representing the set of additive MIMO models, and $\mathcal{S}$ representing the set of structured additive MIMO models, where $\mathcal{S} \subset \mathcal{A}$. We are interested in estimating the parameter vector $\boldsymbol{\rho}$ describing a model in $ \mathcal{S}$. To this end, we assume the existence of a smooth and injective map $\boldsymbol{\mathrm{f}}(\boldsymbol{\rho}) = \boldsymbol{\beta}$ that links the parameter vector $\boldsymbol{\rho}$ of a structured additive MIMO model with the parameter vector $\boldsymbol{\beta}$ of an unstructured additive MIMO model. The Jacobian of such function is assumed to have full column rank for all $\bm{\rho}$, except possibly on a set of Lebesgue measure zero in the parameter space.

The structured additive MIMO model is obtained by solving the following optimization problem
\begin{equation}
\label{ipemoptimization}
    \hat{\bm{\rho}}_N \hspace{-0.04cm}= \hspace{-0.04cm} \underset{\bm{\rho}}{\arg\min} \hspace{0.1cm} V(\bm{\rho}), \hspace{-0.07cm} \quad  V(\bm{\rho}) \hspace{-0.04cm} = \hspace{-0.04cm} \frac{1}{2} \left\|\hat{\bm{\beta}}_N \hspace{-0.04cm}-\hspace{-0.04cm}\mathbf{f}(\bm{\rho})\right\|_{\mathbf{Q}_N^{-1}}^2,
\end{equation}
where $\mathbf{Q}_N\succ \mathbf{0}$ is a weighting matrix, and $\hat{\bm{\beta}}_N$ is the parameter vector obtained from the iterations in \eqref{eq: CT - estimator}.

The method in \eqref{ipemoptimization} ensures the convergence of $\hat{\bm{\rho}}_N$ to the true parameter vector $\bm{\rho}^*$, provided that the sufficient conditions for consistency of $\hat{\bm{\beta}}_N$ specified in Section \ref{sec:statistical} are met. Theorem \ref{thmasymptoticcovarianceipem} presents the asymptotic distribution of $\hat{\bm{\rho}}_N$, and a practical implementation for the optimal selection of the weighting matrix $\mathbf{Q}_N$.

\begin{theorem}[As. dist. structured additive model]
    \label{thmasymptoticcovarianceipem}
    Consider the system output \eqref{eq: CT - output observation}, where the noise $\mathbf{v}(t_k)$ is independent and identically distributed Gaussian white noise with covariance matrix $\bm{\Sigma}_0$. Assume that the iterations in \eqref{eq: CT - estimator} converge for all $N$ sufficiently large, resulting in the estimator $\hat{\bm{\beta}}_N$, and suppose Assumptions \ref{assumption: 1 - CT - consistency} to \ref{assumption: 4 - CT - consistency} hold. Furthermore, assume that the ordering of the submodels described in $\hat{\boldsymbol{\rho}}_N$ in \eqref{ipemoptimization} aligns with $\bm{\rho}^*$. Then, the estimator $\hat{\boldsymbol{\rho}}_N$ is asymptotically Gaussian distributed, i.e.,
\begin{equation}
\label{asymptoticdistributionipem}
\sqrt{N}(\hat{\boldsymbol{\rho}}_N - \boldsymbol{\rho}^*) \xrightarrow{\text {dist.}} \mathcal{N}(\mathbf{0}, \mathbf{P}_{\textnormal{S}}),    
\end{equation}
where
\begin{equation}
\label{pmatrixipem}
    \mathbf{P}_{\textnormal{S}} = (\mathbf{J}^\top_{\mathbf{f}} \mathbf{Q}^{-1} \mathbf{J}_{\mathbf{f}})^{-1} \mathbf{J}^\top_{\mathbf{f}} \mathbf{Q}^{-1} \mathbf{P} \mathbf{Q}^{-1} \mathbf{J}_{\mathbf{f}} (\mathbf{J}^\top_{\mathbf{f}} \mathbf{Q}^{-1} \mathbf{J}_{\mathbf{f}})^{-1},
\end{equation}
with $\mathbf{J}_{\mathbf{f}}$ being the Jacobian matrix of $\mathbf{f}$ evaluated at $\bm{\rho}=\bm{\rho}^*$, and $\mathbf{P}$ being given by \eqref{pmatrix}, for either open or closed-loop scenarios. The asymptotic covariance of $\hat{\bm{\rho}}_N$ is minimized in a positive semidefinite sense if $\mathbf{Q}_N$ is given by the approximate covariance expression in \eqref{eq: CT - estimate asymptotic covariance matrix estimator}. In such case, $ \mathbf{P}_{\textnormal{S}}= (\mathbf{J}^\top_{\mathbf{f}} \mathbf{P}^{-1} \mathbf{J}_{\mathbf{f}})^{-1}$.
\end{theorem}

\textit{Proof.} A Taylor series expansion of $\frac{\partial V(\bm{\rho})}{\partial \bm{\rho}}|_{\bm{\rho}=\hat{\bm{\rho}}_N}$ around the point $\hat{\bm{\rho}}_N = \bm{\rho}^*$ gives
\begin{equation}
    V'(\bm{\rho}^*) + V''(\bm{\rho}^*) (\hat{\bm{\rho}}_N-\bm{\rho}^*) = o_p(\|\hat{\bm{\rho}}_N -\bm{\rho}^*\|),
\end{equation}
where
\begin{align}
    V'(\bm{\rho}^*) &= \mathbf{J}^\top_{\mathbf{f}} \mathbf{Q}^{-1} (\hat{\bm{\beta}}_N -\bm{\beta}^*) + o_p(N^{-1/2}), \notag \\
    V''(\bm{\rho}^*) &= \mathbf{J}^\top_{\mathbf{f}} \mathbf{Q}^{-1} \mathbf{J}_{\mathbf{f}} + o_p(1), \notag
\end{align}
where $\mathbf{J}_{\mathbf{f}}$ is the Jacobian of $\mathbf{f}$ with respect to $\bm{\rho}$, evaluated at $\bm{\rho}= \bm{\rho}^*$. Replacing these matrices above, we find
\begin{equation}
    \sqrt{\hspace{-0.02cm}N}(\hat{\bm{\rho}}_N \hspace{-0.03cm}-\hspace{-0.02cm}\bm{\rho}^*\hspace{-0.01cm}) \hspace{-0.06cm}=\hspace{-0.05cm} (\mathbf{J}^{\hspace{-0.02cm}\top}_{\mathbf{f}} \mathbf{Q}^{\hspace{-0.02cm}-\hspace{-0.02cm}1} \hspace{-0.02cm}\mathbf{J}_{\mathbf{f}})^{\hspace{-0.03cm}-\hspace{-0.02cm}1} \hspace{-0.02cm}\mathbf{J}^{\hspace{-0.02cm}\top}_{\mathbf{f}} \hspace{-0.02cm}\mathbf{Q}^{\hspace{-0.03cm}-\hspace{-0.02cm}1} \hspace{-0.03cm}\sqrt{\hspace{-0.02cm}N}(\hat{\bm{\beta}}_N -\bm{\beta}^*) + o_p(1), \notag
\end{equation}
which, after applying Slutsky's theorem and Theorem \ref{thmefficiency}, leads to \eqref{asymptoticdistributionipem} and \eqref{pmatrixipem}. To minimize \eqref{pmatrixipem} with respect to $\mathbf{Q}$, define $\mathbf{z} = \mathbf{Q}^{-1} \mathbf{J}_{\mathbf{f}} (\mathbf{J}^\top_{\mathbf{f}} \mathbf{Q}^{-1} \mathbf{J}_{\mathbf{f}})^{-1}\mathbf{x}$, where $\mathbf{x}$ is a fixed vector of appropriate dimensions. We have, for any $\mathbf{Q}\succ \mathbf{0}$,
\begin{equation}
    \mathbf{x}^{\hspace{-0.03cm}\top}\hspace{-0.02cm}\mathbf{P}_{\textnormal{S}} \mathbf{x} \hspace{-0.03cm} = \hspace{-0.03cm}\mathbf{z}^{\hspace{-0.03cm}\top}\hspace{-0.02cm} \mathbf{P} \mathbf{z} \geq \hspace{-0.06cm}\min_{\mathbf{z}\in\{\mathbf{z}| \mathbf{J}_{\mathbf{f}}^\top \mathbf{z}=\mathbf{x}\}} \hspace{-0.08cm}\mathbf{z}^{\hspace{-0.03cm}\top}\hspace{-0.02cm} \mathbf{P} \mathbf{z} \hspace{-0.04cm}=\hspace{-0.04cm} \mathbf{x}^{\hspace{-0.04cm}\top}\hspace{-0.06cm} (\mathbf{J}^{\hspace{-0.02cm}\top}_{\mathbf{f}} \hspace{-0.02cm}\mathbf{P}^{-\hspace{-0.02cm}1} \hspace{-0.03cm}\mathbf{J}_{\mathbf{f}})^{\hspace{-0.02cm}-\hspace{-0.02cm}1} \hspace{-0.02cm}\mathbf{x}, \notag 
\end{equation}
where the constrained ordinary least-squares problem above is solved following, e.g., \cite[Section 10.2.2]{gourieroux1995statistics}, with minimizer $\mathbf{z}_{\textnormal{min}}=\mathbf{P}^{-1}\mathbf{J}_{\mathbf{f}} (\mathbf{J}_{\mathbf{f}}^\top \mathbf{P}^{-1}\mathbf{J}_{\mathbf{f}})^{-1} \mathbf{x}$. The minimum is reached when $\mathbf{Q}=\mathbf{P}$, and the selection of $\mathbf{Q}_N$ as \eqref{eq: CT - estimate asymptotic covariance matrix estimator} ensures the convergence to  $\mathbf{P}$ as $N\to\infty$ according to Remark \ref{remarkcovariance}. Thus, the minimum asymptotic covariance is obtained for such selection of $\mathbf{Q}_N$, concluding the proof.  \hfill $\square$

For the open-loop scenario, Theorem \ref{thmasymptoticcovarianceipem} indicates that the method in \eqref{ipemoptimization} has the following asymptotic covariance
\begin{equation}
    \mathbf{P}_{\textnormal{S}}= \overline{\mathbb{E}}\left\{\frac{\partial \boldsymbol{\varepsilon}^\top (t_k,\boldsymbol{\rho})}{\partial \bm{\rho}}\bigg|_{\bm{\rho}=\bm{\rho}^*}\boldsymbol{\Sigma}_0^{-1} \frac{\partial \boldsymbol{\varepsilon}(t_k,\boldsymbol{\rho})}{\partial \bm{\rho}^\top }\bigg|_{\bm{\rho}=\bm{\rho}^*}\right\}^{-1}, \notag 
\end{equation}
where $\bm{\varepsilon}(t_k,\bm{\rho})$ is the residual \eqref{eq: CT - residual function} expressed in terms of the parameter vector of the structured additive MIMO model. This expression coincides with the CRLB for this model structure (c.f. \eqref{fisherinformation}), which means that the proposed two-step approach is asymptotically efficient. Indeed, \eqref{ipemoptimization} is an instance of the indirect prediction error method \cite{Soderstrom1991AnIdentification} for the open-loop scenario, since the refined instrumental variable method in Section \ref{sec:riv} minimizes the prediction error at convergence in iterations and as the sample size tends to infinity \cite{vanderHulst2024additivecdc}.

If the data is obtained from closed-loop operation, then a similar identity to \eqref{matrixinequality} is satisfied, and the proposed method does not reach the CRLB. However, in contrast to the prediction error method, the consistency of the proposed approach in terms of the structured parameters does not require modeling noise dynamics. The two-stage approach can be promising in terms of robustness, as the optimization problem can be interpreted as a curve fitting problem in the parameter space, instead of the data space. As a result, the cost function in \eqref{ipemoptimization} is typically well-behaved, and can be solved efficiently, even if rank constraints on the numerator matrices are enforced.

\begin{rem}
The two step procedure described in this section shares both similarities and differences with the method presented in \cite{Voorhoeve2021IdentifyingStage}. While both approaches enable the identification of modal models, the proposed method i) does not require a preliminary nonparametric identification step to estimate the frequency response function; ii) is applicable in both open loop and closed loop settings, with optimality guarantees in each case, and iii) it extends beyond modal models and can be applied to any structured additive MIMO model structure.  
\end{rem}

\section{Simulation results} \label{sec:simulations}
This section aims to corroborate the theoretical findings in Sections \ref{sec:statistical} and \ref{sec:ipem} through extensive numerical simulations. First, the simulation settings are detailed, and thereafter, Monte Carlo simulations are presented for both the open- and closed-loop scenarios.

\subsection{Simulation settings}
\label{subsec:settings}
\begin{figure}[ht]
    \centering
    \includegraphics[width=0.465\textwidth]{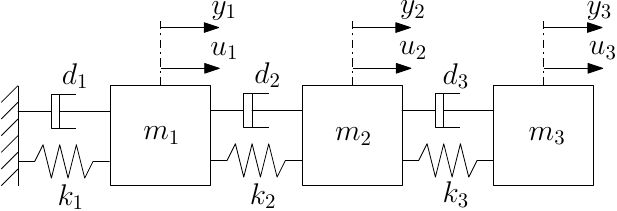}
    \caption{Multivariate simulation model of a flexible motion system with nominal parameter values $m_1 = m_2 =m_3 = 1\ \text{[kg]}$, $k_1 = k_2 = k_3 = 50\ \text{[kN/m]}$ and $d_1 = 0.4955\ \text{[kN/ms]}$, $d_2 = 0.1123\ \text{[kN/ms]}$, and $d_3 =0.1367\ \text{[kN/ms]}$. For this example, the damper coefficients were chosen such that the damping ratios associated to each natural frequency is equal to $0.02$.}
    \label{fig: system simulation example}
\end{figure}

\begin{figure}[ht]
    \centering
    \includegraphics[width=\linewidth]{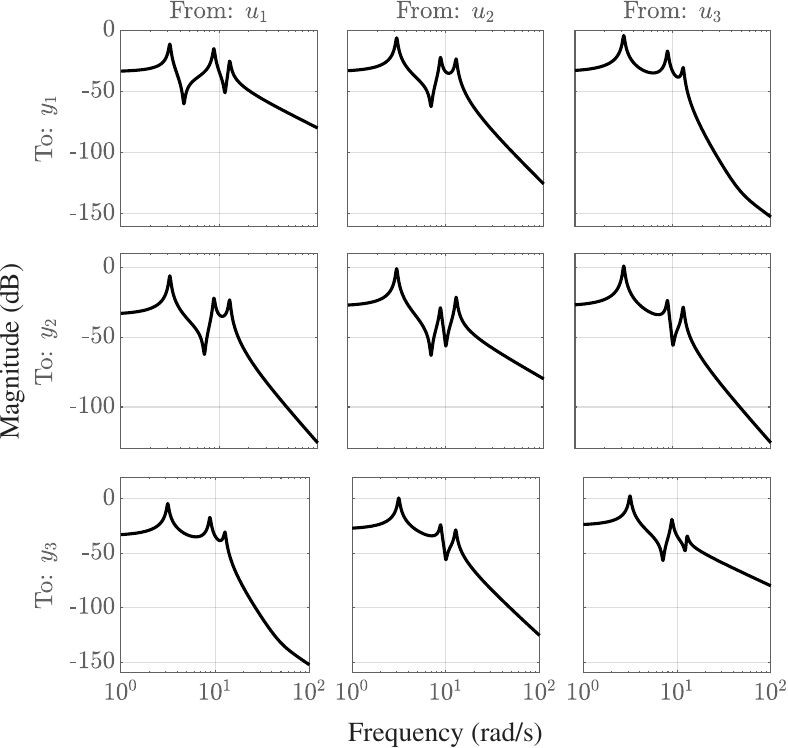}
    \caption{Frequency response of the MIMO system under study.}
    \label{fig: bode system simulation example}
\end{figure}

For the following tests we consider a 3-mass spring damper system with $n_{\textnormal{u}}=3$ inputs and $n_{\textnormal{y}}=3$ outputs, with the free body diagram shown in Figure \ref{fig: system simulation example}, and with Bode plots of the true open-loop MIMO system provided in Figure \ref{fig: bode system simulation example}. This system can be exactly described as an additive MIMO transfer function of the form \eqref{multivariablemodal}, where $K=3$ flexible modes are included. The numerator and denominator parameters can be obtained in terms of $m_i,k_i$ and $d_i$ from the modal modeling approach found in \cite{DeKraker2004ADynamics}.


Two experiments are conducted, where in the first experiment, the system is acting in the open-loop setting, and the second experiment, the system is controlled in closed-loop. The external excitation ($\mathbf{u}(t_k)$ in open loop and $\mathbf{r}(t_k)$ in closed loop) is a Gaussian white signal of unitary variance, and the noise $\mathbf{v}(t_k)$ is given by
\begin{equation}
    \mathbf{v}(t_k) = \frac{1+0.5q^{-1}}{1-0.85q^{-1}}\mathbf{e}(t_k), \notag
\end{equation}
where $\mathbf{e}(t_k)$ is white noise of variance set to yield an output SNR of approximately $30$ [dB] in both settings. The system is controlled in closed loop using a low-bandwidth controller, and is designed according to the loop-shaping rules specified in \cite{oomen2021control}.

We aim to verify the consistency of the refined instrumental variable estimators for these setups, as well as the covariance reduction achieved when estimating a structured additive model. To this end, each experiment considers 50 different sample sizes, ranging logarithmically from $= 10^3$ to $N=10^5$. For each sample size $M= 300$ Monte Carlo runs are performed under different input and noise realizations, and the performance is characterized by the mean squared error (MSE) of six individual parameters: three denominator coefficients $a_{1,2}^*=0.101$, $a_{2,2}^*=0.0129$, $a_{3,2}^*=0.0062$, and three numerator coefficients, corresponding to the $(3,3)$ entry of the $\mathbf{B}_{i,0}^*$ matrix: $\mathbf{B}_{1,0}^{<3,3>}=0.0548$, $\mathbf{B}_{2,0}^{<3,3>}=0.0045$, and $\mathbf{B}_{3,0}^{<3,3>}=0.0007$. If $\hat{\boldsymbol{\beta}}_i$ denotes the $i$th element of the estimated parameter vector, with true value $\boldsymbol{\beta}_i^*$, the MSE is defined as
\begin{equation}
    \text{MSE}(\hat{\boldsymbol{\beta}}_i) = \frac{1}{M} \sum_{i=1}^M \big(\hat{\boldsymbol{\beta}}_i - \boldsymbol{\beta}_i^* \big)^2. \notag
\end{equation}
All estimators are initialized at a random system whose parameters deviate at most $2.5\%$ from their true values, and the assumed intersample behavior for all signals is a ZOH with sampling frequency $f_s = 100$ [Hz].

\subsection{Monte Carlo simulation results}
\label{subsec:mcresults}
\begin{figure}[ht]
    \centering
    \includegraphics[width=0.98\linewidth]{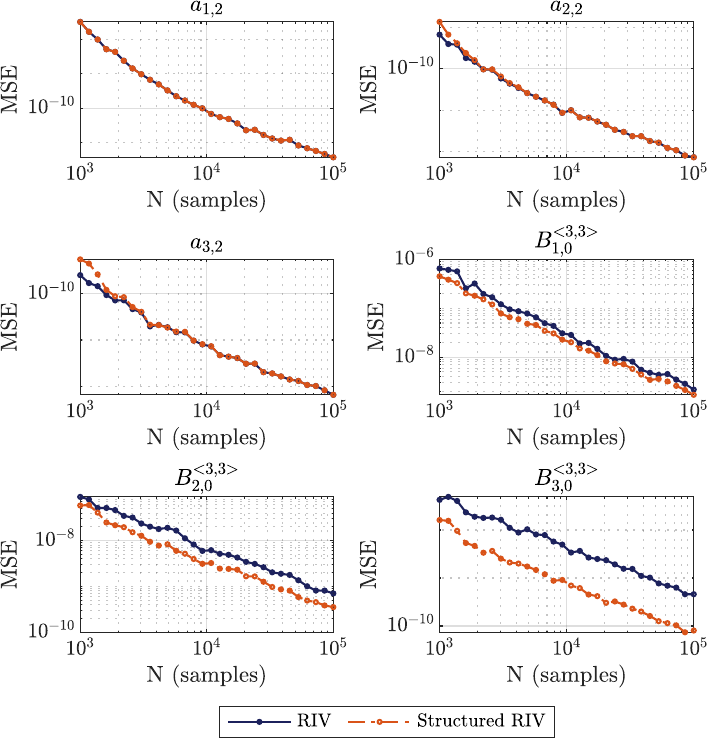}
    \caption{Open-loop setting: MSEs of each estimated parameter with respect to the sample size. The proposed structured RIV method is consistent and delivers improved MSEs for the numerator coefficients.}
    \label{fig: CT - experiment OL}
\end{figure}

We begin by presenting the open-loop system simulation results. In Figure \ref{fig: CT - experiment OL}, we compare the MSEs of the chosen denominator and numerator coefficients for the refined instrumental variable method for MIMO additive systems and the structured additive MIMO presented in Section \ref{sec:ipem}. The MSEs of all parameters decay to zero as more samples are considered, supporting the consistency of both estimators \cite{lehmann1998theory}, aligning  with Theorem \ref{theorem: CT - consistency}. Moreover, the rank-1 enforcement in the numerator matrices leads to a variance reduction of the estimated numerator parameters, which is in agreement with Theorem \ref{thmasymptoticcovarianceipem}. 

\begin{figure}[ht]
    \centering
    \includegraphics[width=0.98\linewidth]{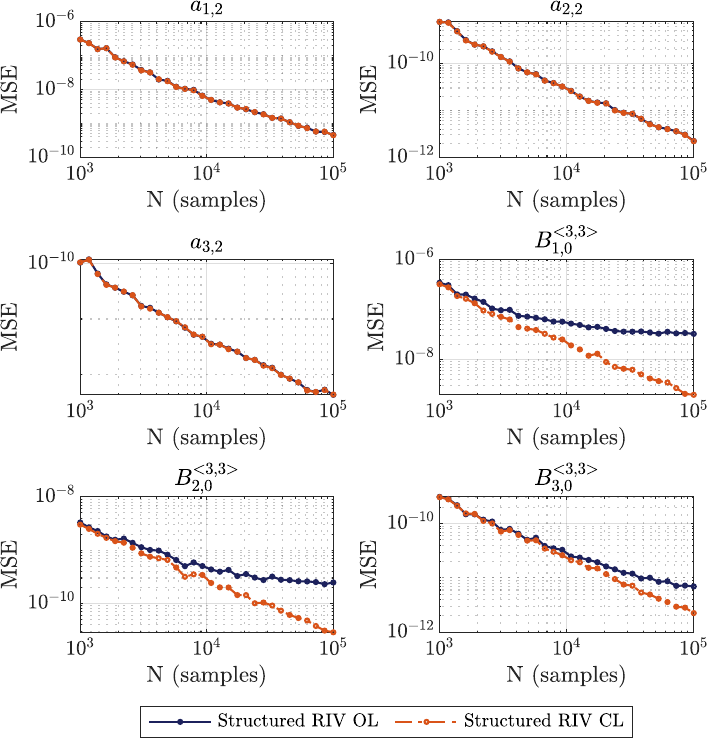}
    \caption{Closed-loop setting: MSEs of each estimated parameter with respect to the sample size. The structured RIV method for closed-loop systems is consistent, while bias is observed if the the open-loop method is used.}
    \label{fig: CT - experiment CL}
\end{figure}

Additionally, we present the simulation results of the closed-loop experiment in Figure \ref{fig: CT - experiment CL}. For this experiment, we compare the open and closed-loop variants of the method in Section \ref{sec:ipem} for structured additive MIMO system identification. The open-loop method only uses the input and output information $\{\mathbf{u}(t_k),\mathbf{y}(t_k)\}_{k=1}^N$, while the closed-loop variant also considers the reference signal $\{\mathbf{r}(t_k)\}_{k=1}^N$. The results confirm that the closed-loop identification method for structured additive MIMO systems is consistent even if the noise model is not estimated, in alignment with Theorem \ref{theorem: CT - consistency}. This is not the case for the open-loop variant, which is shown to be not consistent in Figure \ref{fig: CT - experiment CL} as the MSE curves pertaining to the numerator terms do not decrease to zero as the number of samples increases. In this scenario, the open-loop variant would require estimating the noise model accurately mitigate the asymptotic bias \cite{forssell1999closed}, a condition that is not required for the proposed closed-loop variant.

\section{Conclusions}
\label{sec:conclusions}
In this paper we presented and analyzed a time-domain method that enables the identification of structured additive continuous-time MIMO systems with consistency guarantees. The structured model is obtained in two steps. First, an additive MIMO system identification method with unconstrained numerator polynomials was formulated as a solution of correlation equations. These equations differ only in the instrument matrix selection for open- and closed-loop settings, and they were solved via refined instrumental variables. A comprehensive statistical analysis established that the estimator is generically consistent and achieves minimum asymptotic covariance among consistent estimators of this model structure. In a second step, the asymptotic covariance expressions were leveraged to optimize for a structured additive model, further reducing asymptotic covariance and leading to more parsimonious MIMO models. Extensive simulations were performed to validate the theoretical findings, and to show the efficacy of the proposed approaches.
\appendix 	
\section{Technical lemmas}
\label{appendixA}
\begin{lemma} Under Assumptions \ref{assumption1} to \ref{assumption4}, the matrix $\mathbf{\Psi}_\mathbf{z}$ in (\ref{eq: CT - consistency - psy}) is positive definite when evaluated at $\bm{\beta}=\bm{\beta}^*$.
\label{lemma: pos def}
\end{lemma}
\textit{Proof}: Denote the matrix in \eqref{eq: CT - consistency - psy} evaluated at the true system parameters as $\mathbf{\Psi}_\mathbf{z}^*$. The matrix $\mathbf{\Psi}_\mathbf{z}^*$ is then positive definite if  $\mathbf{x}^\top\mathbf{\Psi}_\mathbf{z}^*\mathbf{x} = 0$ implies $\mathbf{x} = \mathbf{0}$. Substituting $\bm{\mathcal{Z}}_i(t_k) = \mathbf{M}_i(p)\mathbf{Z}(t_k)$ in $\mathbf{\Psi}_\mathbf{z}^*$ allows us to write 
\begin{equation}
    \mathbf{x}^\top\mathbf{\Psi}_\mathbf{z}^*\mathbf{x} = \overline{\mathbb{E}}\left\{ \left\| \mathbf{Z}^\top(t_k) \sum\limits_{i=1}^K \frac{\mathbf{M}^\top_i(p)\mathbf{x}_i}{{A^*_i}^2(p)} \right\|_{\bm{\Sigma}_0^{-1}}^2\right\}, \notag
\end{equation}
where we have partitioned $\mathbf{x}$ as $\mathbf{x}=[\mathbf{x}^\top_1,\dots, \mathbf{x}_K^\top]^\top$. Combining the fractions under a common denominator yields
\begin{align}
\mathbf{x}^{\hspace{-0.05cm}\top}\hspace{-0.07cm}\mathbf{\Psi}_{\hspace{-0.02cm}\mathbf{z}}^*\mathbf{x} \hspace{-0.07cm}=\hspace{-0.07cm} \overline{\mathbb{E}}\hspace{-0.02cm}\Biggl\{\hspace{-0.03cm} \Biggl \|   \frac{\mathbf{Z}^\top(t_k)}{\prod_{i\hspace{-0.02cm}=\hspace{-0.02cm}1}^K\hspace{-0.08cm}{A_i^*}^2\hspace{-0.03cm}(p)} \hspace{-0.12cm}\sum\limits_{i=1}^K \hspace{-0.05cm}\mathbf{M}^\top_i\hspace{-0.06cm}(p)\mathbf{x}_i\hspace{-0.2cm} \prod\limits_{\substack{j=1, \ldots, K \\ j \neq i}}\hspace{-0.23cm}{A_j^*}^2\hspace{-0.04cm}(p) \Biggr\|_{\hspace{-0.02cm}\bm{\Sigma}_0^{\hspace{-0.03cm}-\hspace{-0.03cm}1}}^2\hspace{-0.08cm}\Biggr\}\hspace{-0.03cm}. \notag
\end{align}
After introducing the $(n_{\textnormal{y}}\times n_{\textnormal{u}})$-dimensional filters $\mathbf{Q}_i(p)=\operatorname{vec}^{-1} \{\mathbf{M}^\top_i(p)\mathbf{x}_i\}$, the following identity holds:
\begin{equation}
    \mathbf{Z}^{\hspace{-0.02cm}\top}\hspace{-0.04cm}(t_k) \mathbf{M}^{\hspace{-0.02cm}\top}_i\hspace{-0.05cm}(p)\mathbf{x}_i\hspace{-0.07cm}=\hspace{-0.07cm} \left(\mathbf{z}^{\hspace{-0.03cm}\top}\hspace{-0.04cm}(t_k)\hspace{-0.03cm}\otimes\hspace{-0.03cm} \mathbf{I}_{n_{\textnormal{y}}}\right)\hspace{-0.05cm}\operatorname{vec}(\mathbf{Q}_i(p))\hspace{-0.07cm}=\hspace{-0.07cm}  \mathbf{Q}_i(p)\mathbf{z}(t_k), \notag
\end{equation}
leading to
\begin{equation}
\label{parsevaltimedomain}
    \mathbf{x}^\top\mathbf{\Psi}_\mathbf{z}^*\mathbf{x} = \overline{\mathbb{E}}\left\{\left\| \frac{\mathbf{B}_{\mathbf{x}}(p)}{\prod_{i=1}^K{A_i^*}^2(p)} \mathbf{z}(t_k) \right\|^2_{\bm{\Sigma}^{-1}_0} \right\},
\end{equation}
with the $(n_{\textnormal{y}}\times n_{\textnormal{u}})$-dimensional matrix polynomial  $\mathbf{B}_{\mathbf{x}}(p)$ being given by
\begin{equation}
\mathbf{B}_{\mathbf{x}}(p) =  \sum\limits_{i=1}^K \mathbf{Q}_i(p) \prod\limits_{\substack{j=1, \ldots, K \\ j \neq i}}{A_j^*}^2(p).
\end{equation}
Using Parseval's theorem \cite[p. 61]{soderstrom2012discrete}, the expression in \eqref{parsevaltimedomain} is written in the frequency domain as
\begin{equation}
\mathbf{x}^\top\mathbf{\Psi}_\mathbf{z}^*\mathbf{x} \hspace{-0.05cm}=\hspace{-0.05cm} \frac{1}{2 \pi} \int_{-\pi}^\pi \operatorname{Tr}\left\{ \bm{\Sigma}_0^{-1}\tilde{\mathbf{B}}_{\mathbf{x}}(e^{\mathrm{i}\omega})  \tilde{\boldsymbol{\Phi}}_{\mathbf{z}}(\omega) \tilde{\mathbf{B}}^\hop_{\mathbf{x}}(e^{\mathrm{i}\omega})\right\}\mathrm{d}\omega, \notag 
\end{equation}
where $\tilde{\boldsymbol{\Phi}}_{\mathbf{z}}(\omega)$ is the spectral distribution of $\mathbf{z}(t_k)$ filtered by $1/ \prod_{i=1}^K\tilde{A}_{i}^{*2}(e^{\mathrm{i} \omega})$, with $\tilde{A}_{i}^{*}(e^{\mathrm{i}\omega})$ and $\tilde{\mathbf{B}}_{\mathbf{x}}(e^{\mathrm{i}\omega})$ being the ZOH discrete-time denominator and numerator filters of $\mathbf{B}_{\mathbf{x}}(p)/\prod_{i=1}^K{A_i^*}^2(p)$ evaluated at $e^{\mathrm{i}\omega}$. Note that $\tilde{\boldsymbol{\Phi}}_{\mathbf{z}}(\omega)$ consists of at least $2\sum_{i=1}^K n_i$ spectral lines, since filtering the input signal $\mathbf{z}(t_k)$ by an asymptotically stable filter does not affect the persistence of excitation order. Therefore, $\mathbf{x}^\top\mathbf{\Psi}_\mathbf{z}^*\mathbf{x} = 0$ implies $ \operatorname{Tr}\{\bm{\Sigma}_0^{-1}\tilde{\mathbf{B}}_{\mathbf{x}}(e^{\mathrm{i} \omega})  \tilde{\boldsymbol{\Phi}}_{\mathbf{z}}(\omega) \tilde{\mathbf{B}}^\hop_{\mathbf{x}}(e^{\mathrm{i} \omega})\}=0$ for almost all $\omega$, i.e.,
\begin{equation}
    \tilde{\mathbf{B}}_{\mathbf{x},j}^\hop(e^{\mathrm{i} \omega})  \tilde{\boldsymbol{\Phi}}_{\mathbf{z}}(\omega) \tilde{\mathbf{B}}_{\mathbf{x},j}(e^{\mathrm{i} \omega}) = \mathbf{0},\quad j=1,\dots,n_{\textnormal{y}},\notag 
\end{equation}
where $\mathbf{B}_{\mathbf{x},j}(e^{\mathrm{i}\omega})$ denotes the $j$th column of the matrix $\mathbf{B}_{\mathbf{x}}^\hop(e^{\mathrm{i}\omega})$. Since each vector polynomial $\mathbf{B}_{\mathbf{x},j}(e^{\mathrm{i}\omega})$ is of order at most $2\sum_{i=1}^K n_i$, the definition of persistently exciting signals leads to $\tilde{\mathbf{B}}_{\mathbf{x}}(e^{\mathrm{i} \omega}) \equiv \mathbf{0}$. From Assumption \ref{assumption4}, we find that the discrete-time equivalent is unique \cite{kollar1996equivalence}, which implies that ${\mathbf{B}}_{\mathbf{x}}(p) \equiv \mathbf{0}$. Consequently, ${\mathbf{B}}_{\mathbf{x}}(p) \equiv \mathbf{0}$ implies that ${\mathbf{Q}}_{i}(p) \equiv \mathbf{0}$ for $i = 1,\ldots,K$ by applying the same reasoning outlined in Lemma 3 of \cite{Gonzalez2024IdentificationClosed-loop}. This result establishes that $\mathbf{x} = \mathbf{0}$, which in turn ensures that the matrix $\mathbf{\Psi}_\mathbf{z}^*$ is positive definite. \hfill $\square$

\begin{lemma}
\label{lemmaqss}
If $\hat{\bm{\beta}}_N\to \bm{\beta}^*$ with probability 1 as $N\to \infty$, then with probability 1,
\begin{equation}
    \frac{1}{N}\hspace{-0.05cm}\sum_{k=1}^N \hspace{-0.04cm}\bm{\varepsilon}(t_k,\hat{\bm{\beta}}_N)\bm{\varepsilon}^\top(t_k,\hat{\bm{\beta}}_N) \hspace{-0.06cm}\to\hspace{-0.06cm} \bm{\Sigma}_0 \textnormal{ as }N\to \infty.
\end{equation}
\end{lemma}
\textit{Proof}: Denote by $\bm{\Sigma}(\bm{\beta}^*)$ the covariance matrix estimate \eqref{eq: CT - ML noise covariance} evaluated at the true parameters, and let $\epsilon>0$. Since $\hat{\bm{\beta}}_N\to \bm{\beta}^*$ with probability 1 as $N\to \infty$, we can pick an integer $N_\epsilon$ such that
\begin{equation}
    \sup_{k\geq N_\epsilon}\|\bm{\varepsilon}(t_k,\hat{\bm{\beta}}_N)\bm{\varepsilon}^\top(t_k,\hat{\bm{\beta}}_N)-\bm{\varepsilon}(t_k,\bm{\beta}^*)\bm{\varepsilon}^\top(t_k,\bm{\beta}^*)\|<\epsilon. \notag 
\end{equation}
Hence, for $N$ large enough, the following chain of inequalities holds with probability 1:
\begin{align}
&\|\bm{\Sigma}(\hat{\bm{\beta}}_N)-\bm{\Sigma}(\bm{\beta}^*)\|\leq \frac{(N-N_\epsilon)\epsilon}{N} \notag \\
&+\hspace{-0.07cm}\frac{1}{N} \hspace{-0.06cm}\sum_{k=1}^{N_\epsilon} \hspace{-0.05cm}\left\|\bm{\varepsilon}(t_k,\hspace{-0.02cm}\hat{\bm{\beta}}_N\hspace{-0.01cm})\bm{\varepsilon}^{\hspace{-0.03cm}\top}\hspace{-0.05cm}(t_k,\hspace{-0.02cm}\hat{\bm{\beta}}_N)\hspace{-0.05cm}-\hspace{-0.02cm}\bm{\varepsilon}(t_k,\hspace{-0.02cm}\bm{\beta}^*)\bm{\varepsilon}^{\hspace{-0.03cm}\top}\hspace{-0.05cm}(t_k,\hspace{-0.02cm}\bm{\beta}^*) \right\|\hspace{-0.07cm} <\hspace{-0.04cm}2\epsilon. \notag
\end{align}
Since $\epsilon$ is arbitrary, we conclude that $\|\bm{\Sigma}(\hat{\bm{\beta}}_N)-\bm{\Sigma}(\bm{\beta}^*)\|=o_p(1)$. Thus, the triangle inequality yields
\begin{equation}
    \|\bm{\Sigma}(\hat{\bm{\beta}}_N\hspace{-0.01cm})\hspace{-0.04cm}-\hspace{-0.03cm}\bm{\Sigma}_0\|\hspace{-0.06cm}\leq \hspace{-0.06cm}\|\bm{\Sigma}(\hat{\bm{\beta}}_N\hspace{-0.01cm})\hspace{-0.04cm}-\hspace{-0.03cm}\bm{\Sigma}(\bm{\beta}^*\hspace{-0.01cm})\|\hspace{-0.04cm} +\hspace{-0.04cm} \|\bm{\Sigma}(\bm{\beta}^*)\hspace{-0.04cm}-\hspace{-0.03cm}\bm{\Sigma}_0\|\hspace{-0.06cm}=\hspace{-0.06cm} o_p\hspace{-0.02cm}(1), \notag 
\end{equation}
where $ \|\bm{\Sigma}(\bm{\beta}^*)-\bm{\Sigma}_0\|=o_p(1)$ holds due to Theorem 2.3 of \cite{Ljung1999SystemUser}. \hfill $\square$

\begin{lemma}
\label{lemmaMv}
Under the same assumptions stated in Theorem \ref{thmefficiency},
\begin{equation}
\label{lemma3statement}
\sum_{k=1}^N \frac{\partial}{\partial \bm{\beta}^{\top}}[\hat{\boldsymbol{\Phi}}(t_k,\boldsymbol{\beta}) \boldsymbol{\Sigma}^{-1}(\bm{\beta})]\bigg|_{\bm{\beta}=\bm{\beta}^*} \mathbf{v}(t_k)=o_p(N).
\end{equation}
\end{lemma}
\textit{Proof}: Denote $\mathbf{M}_i(t_k)$ as the partial derivative of $\hat{\boldsymbol{\Phi}}(t_k,\boldsymbol{\beta}) \boldsymbol{\Sigma}^{-1}(\bm{\beta})$ with respect to the $i$th entry of $\bm{\beta}$, i.e.,
\begin{align}
    \label{Mi}
    \mathbf{M}_i\hspace{-0.02cm}(\hspace{-0.01cm}t_k\hspace{-0.01cm})\hspace{-0.08cm}= \hspace{-0.08cm}\frac{\partial \hat{\boldsymbol{\Phi}}\hspace{-0.02cm}(\hspace{-0.01cm}t_k,\hspace{-0.02cm}\boldsymbol{\beta})}{\partial \bm{\beta}_i}\hspace{-0.02cm} \bigg|_{\hspace{-0.02cm}\bm{\beta}\hspace{-0.02cm}=\hspace{-0.02cm}\bm{\beta}^*}\hspace{-0.28cm}\bm{\Sigma}_*^{-\hspace{-0.02cm}1}\hspace{-0.12cm} - \hspace{-0.07cm} \hat{\boldsymbol{\Phi}}\hspace{-0.02cm}(t_k,\hspace{-0.02cm}\boldsymbol{\beta}^*\hspace{-0.02cm}) \bm{\Sigma}_*^{-\hspace{-0.02cm}1}\hspace{-0.03cm} \bm{\Sigma}_*' \bm{\Sigma}_*^{-\hspace{-0.02cm}1}\hspace{-0.06cm}, \hspace{-0.14cm}
\end{align}
where we have introduced the short notation $\bm{\Sigma}_* := \bm{\Sigma}(\bm{\beta}^*)$ and $\bm{\Sigma}_{*i}':=\partial \bm{\Sigma}(\bm{\beta})/\partial \bm{\beta}_i\big|_{\bm{\beta}=\bm{\beta}^*}$. The approach for proving \eqref{lemma3statement} consists of showing that $\sum_{k=1}^N \mathbf{M}_i(t_k)\mathbf{v}(t_k)$ is of order $o_p(N)$ for any $i$. Since the partial derivative of the instrument matrix consists of filtered external signals (input or reference) that are independent of the disturbance $\mathbf{v}(t_k)$ according to Assumption \ref{assumption2}, we obtain
\begin{equation}
\label{alsoimplies}
\sum_{k=1}^N \frac{\partial \hat{\boldsymbol{\Phi}}(t_k,\boldsymbol{\beta})}{\partial \bm{\beta}_i} \bigg|_{\bm{\beta}=\bm{\beta}^*}\boldsymbol{\Sigma}_0^{-1} \mathbf{v}(t_k) = o_p(N).
\end{equation}
Using Theorem 2.3 of \cite{Ljung1999SystemUser}, we have $\bm{\Sigma}_0^{-1}-\bm{\Sigma}_*^{-1}=o_p(1)$. Additionally, since the filtered versions of $\mathbf{u}(t_k), \mathbf{r}(t_k)$ and $\mathbf{v}(t_k)$ have bounded second moments, the sum of the norms of each filtered signal is of order $O_p(N)$. Thus, \eqref{alsoimplies} further implies
\begin{equation}
\label{alsoimplies2}
\sum_{k=1}^N \frac{\partial \hat{\boldsymbol{\Phi}}(t_k,\boldsymbol{\beta})}{\partial \bm{\beta}_i} \bigg|_{\bm{\beta}=\bm{\beta}^*}\boldsymbol{\Sigma}_*^{-1}\mathbf{v}(t_k) = o_p(N). 
\end{equation}
We now focus on the second term in \eqref{Mi}, and consider the decomposition
\begin{align}
    &\bm{\Sigma}_*^{-1} \bm{\Sigma}_{*i}' \bm{\Sigma}_*^{-1} = \bm{\Sigma}_0^{-1}\bm{\Sigma}_{*i}' \bm{\Sigma}_0^{-1}+\bm{\Sigma}_0^{-1}\bm{\Sigma}_{*i}' (\bm{\Sigma}_*^{-1}-\bm{\Sigma}_0^{-1}) \notag \\
    &+(\bm{\Sigma}_*^{-1}\hspace{-0.06cm}-\hspace{-0.05cm}\bm{\Sigma}_0^{-1})\bm{\Sigma}_{*i}' \bm{\Sigma}_0^{-1} \hspace{-0.04cm}+ \hspace{-0.03cm}(\bm{\Sigma}_*^{-1}\hspace{-0.06cm}-\hspace{-0.05cm}\bm{\Sigma}_0^{-1})\bm{\Sigma}_{*i}'(\bm{\Sigma}_*^{-1}\hspace{-0.06cm}-\hspace{-0.05cm}\bm{\Sigma}_0^{-1}), \notag
\end{align}
where we notice that all right-hand side terms except for $\bm{\Sigma}_0^{-1}\bm{\Sigma}_{*i}' \bm{\Sigma}_0^{-1}$ are of order $o_p(1)$ due to Theorem 2.3 of \cite{Ljung1999SystemUser}. Similar to the derivation of \eqref{alsoimplies2}, we may replace $\bm{\Sigma}_*^{-1} \bm{\Sigma}_{*i}' \bm{\Sigma}_*^{-1}$ by $\bm{\Sigma}_0^{-1}\overline{\bm{\Sigma}}_{*i}' \bm{\Sigma}_0^{-1}$ by neglecting a factor $o_p(N)$ in the computation of $\sum_{k=1}^N \mathbf{M}_i(t_k)\mathbf{v}(t_k)$, where
\begin{equation}
    \overline{\bm{\Sigma}}_{*i}'= \lim_{N\to\infty} -\frac{1}{N} \sum_{s=1}^N \big( \mathbf{v}(t_s) \mathbf{y}_{*i}'^\top(t_s)+ \mathbf{y}_{*i}'(t_s)\mathbf{v}^\top(t_s) \big), \notag 
\end{equation}
with $\mathbf{y}_{*i}'(t_s) = \partial \mathbf{y}(t_s,\bm{\beta})/\partial \bm{\beta}_i\big|_{\bm{\beta}=\bm{\beta}^*}$, where $\mathbf{y}(t_s,\bm{\beta})$ is the output of the model with parameter vector $\bm{\beta}$. The matrix $\overline{\bm{\Sigma}}_{*i}'=\mathbf{0}$ for open-loop operation, whereas it is finite-valued for closed-loop operation. In any case, we conclude
\begin{equation}
\sum_{k=1}^N \hat{\boldsymbol{\Phi}}(t_k,\boldsymbol{\beta}^*) \bm{\Sigma}_0^{-1} \overline{\bm{\Sigma}}_{*i}' \bm{\Sigma}_0^{-1} \mathbf{v}(t_k) = o_p(N), \notag
\end{equation}
leading to \begin{equation}
\sum_{k=1}^N \hat{\boldsymbol{\Phi}}(t_k,\boldsymbol{\beta}^*) \bm{\Sigma}_0^{-1} \bm{\Sigma}_{*i}' \bm{\Sigma}_0^{-1} \mathbf{v}(t_k) = o_p(N), \notag
\end{equation}
which leads to the desired conclusion.  \hfill $\square$

\bibliography{References}	         
\end{document}